# A paradigm for developing earthquake probability forecasts based on geoelectric data

Authors:

Hong-Jia Chen[1], Chien-Chih Chen[1,4], Guy Ouillon[2], Didier Sornette[3]

Affiliations:

[1] Department of Earth Sciences, National Central University, Zhongli, 32001, Taiwan.

[2] Lithophyse, Nice, France.

[3] Department of Management, Technology and Economics, ETH Zürich, Zürich, Switzerland.

[4] Earthquake-Disaster & Risk Evaluation and Management Center, National Central University, Zhongli, 32001, Taiwan.

Corresponding author: Hong-Jia Chen (redhouse6341@g.ncu.edu.tw)

# Abstract

We examine the precursory behavior of geoelectric signals before large earthquakes by means of an algorithm including an alarm-based model and binary classification. This algorithm, introduced originally by Chen and Chen [*Nat. Hazards.*, **84**, 2016], is improved by removing a time parameter for coarse-graining of earthquake occurrences, as well as by extending the single station method into a joint stations method. We also determine the optimal frequency bands of earthquake-related geoelectric signals with highest signal-to-noise ratio. Using significance tests, we also provide evidence of an underlying seismoelectric relationship. It is appropriate for machine learning to extract this underlying relationship, which could be used to quantify probabilistic forecasts of impending earthquakes, and to get closer to operational earthquake prediction.

# 1. Introduction

Large earthquakes, together with other hazards they trigger, are the deadliest of all natural disasters, killing up to hundreds of thousands of people and causing sizable economic losses. If an impending earthquake could be predicted several days to weeks before it occurs, appropriate measures could be taken to save lives and reduce losses. The progress in earthquake forecasting is hampered by three main constraints: (i) no possibility of direct, in-situ observations, (ii) scarcity of large earthquakes, and (iii) complexity of physical processes of earthquake nucleation. For instance, because crustal stress states are hardly measurable in regional scales, the failure of rocks is difficult to predict so that earthquake forecasting cannot progress as fast as, say, weather forecasting.

Over the past few decades, a number of scientists have been focusing on seismo-electromagnetic precursors in natural settings (see, e.g., Chen et al., 2017; Chen & Chen, 2016; Eftaxias, 2010; Eftaxias et al., 2001; Han et al., 2011; Han et al., 2015, 2017, Varotsos et al., 2002, 2006, 2009 ; see also Freund et al., this volume, for a review). On the other hand, experimental results of rock fracturing tests also support the feasibility of forecasting failure based on seismo-electromagnetic precursors. The possible following models, among others, have been proposed: peroxy-defects theory (Freund, 2003, 2007a, 2007b), piezoelectrics (Finkelstein et al., 1973; Nitsan, 1977; Ogawa et al., 1985; Sornette & Sornette, 1990), piezomagnetism (Revol et al., 1977), stress-induced currents (Hadjicontis et al., 2004; Hadjicontis & Mavromatou, 1994; Mavromatou et al., 2004).

The work reported in this paper evaluates the significance of some of those precursors using field observation data. Chen and Chen (2016) and Chen et al. (2017) have tested statistically the existence of geoelectric precursors associated with moderate to large earthquakes. They have built up a Geoelectric Monitoring System Time of Increased Probability (GEMSTIP) algorithm, testing possible relationships between geoelectric field anomalies and earthquake occurrences. Previous studies (e.g., Chen et al., 2017; Chen & Chen, 2016; Han et al., 2017) optimized the parameters of their predictive models on training datasets only, but did not validate them in forecasting experiments, while a forecasting strategy is claimed practical if and only if the model performances both in the training and forecasting phases are comparable and significant. In addition to reviewing the GEMSTIP algorithm, we improve it by removing one operational time parameter (for coarse-graining of earthquake occurrences), as well as by introducing a joint stations method. Using the new version of the GEMSTIP algorithm, we also determine the frequency bands of earthquake-related geoelectric signals with high signal-to-noise ratio. The investigation of this study thus reveals, from the viewpoint of field observations, if transient anomalous geoelectric fields are generated prior to large earthquakes in a fault zone.

The organization of this paper is as follows. In Section 2, we first describe the data sources of both geoelectric signals and earthquake catalogs. Section 3 illustrates the main parameters influencing the determination of geoelectric anomalies and target earthquakes. Section 4 presents the improved GEMSTIP algorithm for testing relationships between geoelectric field statistics and earthquake occurrences. Section 5 determines the frequency bands of earthquake-related signals by analyzing a great number of high- and low-pass filtered datasets with different cut-off frequencies. Finally, Section 6 presents our conclusions and suggests future works.

## 2. Data description
### 2.1 Geoelectric data

Since 2012, several geoelectric stations have been installed and uniformly distributed on the Taiwan Island (Fig. 1), with a mesh size of approximately 50 km. This network is called the GeoElectric Monitoring System (GEMS). Each station continuously records self-potentials (referred to as geoelectric fields hereafter), which are naturally occurring electric potential differences in the Earth, i.e. passive sources. The geoelectric fields are measured by a non-polarized electrode, buried at about 1-2m depth, relative to a fixed reference one. Each site features two horizontal components, together with GPS synchronization. Information about the geoelectric stations are listed in Table 1, including station onset time, station location, dipole length, and dipole azimuth. The dipole length of each component is within hundreds of meters to a few kilometers. Due to site-dependent limitations, the azimuth of each component is not exactly North or East. The collected data are digitized with 24 bit A/D converters and fed to the data acquisition PC and transferred to Taiwan's Central Weather Bureau (CWB). The management of the GEMS network was transferred from Prof. Chien-Chih Chen's Lab to the CWB in February, 2017. The accuracy of the measured voltages reaches 1 µV, while the sampling rate is 15 Hz.

## 2.2 Earthquake catalog

The network of seismographic stations of Taiwan is the densest in the world and provides a set of abundant waveform data. The CWB has routinely processed seismic waveforms and cataloged the source parameters of all recorded earthquakes, including occurrence time, location, magnitude, etc. Due to the abundance of earthquakes, the seismotectonics of Taiwan is well known (see Kuo-Chen et al., 2012; Tsai, 1986). Because $M_L \geq 5$ earthquakes usually lead to regional disasters, we focus only on $M_L \geq 5$ earthquakes during 2012/01/01 to 2016/12/31 (with the date format of yyyy/mm/dd in UTC+8 hereafter), within the region of 119.5-122.5°E and 21.5-25.5°N, and at all depths, which amounts to 105 events. The distribution of the selected earthquakes is also shown in Fig. 1. Among those events, the strongest is the $M_L$=6.62 earthquake occurred on 2012/06/10 at 122.31°E, 24.46°N (and its focal solution can be found in http://rmt.earth.sinica.edu.tw/). Furthermore, there are five inland earthquakes with $M_L$>6: (i) $M_L$=6.35, 2012/02/26, 120.75°E, 22.75°N (see Chen et al., 2013), (ii) $M_L$=6.24, 2013/03/27, 121.05°E, 23.90°N (see Chuang et al., 2013), (iii) $M_L$=6.48, 2013/06/02, 120.97°E, 23.86°N (see Chuang et al., 2013), (iv) $M_L$=6.42, 2013/10/31, 121.35°E, 23.57°N (see Lee et al., 2014), and (v) $M_L$=6.60, 2016/02/06, 120.54°E, 22.92°N (see Lee et al., 2016). The reported spatial location uncertainty for the 105 selected events is, in average, 0.21 km horizontally and 0.23 km vertically.

## 3. Detection of geoelectric anomalies before earthquakes

### 3.1 Statistical indices of geoelectric fields

In this study, we first resample the geoelectric data of the N- and E-components using a sampling rate of 1 Hz, in the spirit of Eftaxias et al. (2003), Varotsos et al. (2003), and Han et al. (2017) for example.

We then calculate the daily mean ($\mu$), variance ($V$), skewness ($S$), and kurtosis ($K$) of the N- and E-components of the geoelectric fields, as follows:

$$\mu = \frac{1}{N}\sum_{i=1}^{N} x_i,$$

$$V = \frac{1}{N-1}\sum_{i=1}^{N}|x_i - \mu|^2,$$

$$S = \frac{\frac{1}{N}\sum_{i=1}^{N}(x_i-\mu)^3}{\left(\sqrt{\frac{1}{N}\sum_{i=1}^{N}(x_i-\mu)^2}\right)^3},$$

$$K = \frac{\frac{1}{N}\sum_{i=1}^{N}(x_i-\mu)^4}{\left(\frac{1}{N}\sum_{i=1}^{N}(x_i-\mu)^2\right)^2}, \tag{1}$$

where $x_i$ is one component of the geoelectric field, and $i$ spans from $1^{st}$ to $86400^{th}$ sec for each day. Figure 2 shows the time series of the four indices of the N- and E-components for the PULI station. $M_L$≥5 earthquakes occurring within 60 km to this station are also plotted.

The daily mean series in Fig. 2 shows smoother variations, so that it is hard to determine what anomalies are prior to a large earthquake. Furthermore, Chen and Chen (2016) and Chen et al. (2017) indicate that the skewness and kurtosis of geoelectric fields might be affected in an earthquake preparation process. Hence, we focus on the time series of skewness and kurtosis. For simplicity, we consider the absolute values of skewness (|S|).

## 3.2 Definition of geoelectric fields anomalies using skewness and kurtosis

Because the skewness and kurtosis of geoelectric fields are not normally distributed variables, we use the median and interquartile range (IQR) to describe these variables. Both the modulus of skewness (|S|) and kurtosis (K) are larger than or equal to zero. It is clear that there must be an upper threshold above which a value of such a variable is anomalous. Given the time series of a statistical index (y), the upper threshold (θ) is defined as:

$$\theta(t;y) = Median(y_i) + A_{thr} * IQR(y_i), \quad i = t - \Delta t \text{ to } t, \qquad (2)$$

where $t$ is the time in days, $y_i$ is |S| or K within the time interval $\Delta t$ days, and $A_{thr}$ is a factor to tune the level of the upper threshold. Naturally, |S| or K is then defined as anomalous whenever $|S(t)| > \theta(t; |S|)$ or $K(t) > \theta(t; K)$.

In a sensitivity analysis, we test the effect of the choice of $\Delta t$ on calculating the median and IQR of |S| and K. We consider values of $\Delta t$ from 100 to 1200 days, with a step of 100 days. In this paper, ranges of parameters are set explicitly by using the format {start:increment:end}, thus $\Delta t \in$ {100:100:1200} day. For a given station, from

its onset time, we compute the medians and *IQRs* of |*S*| and *K* within intervals of length *Δt* by shifting the time windows by a lag of 0.05*Δt*. We compute the statistics (mean ± 1 standard deviation) of those medians and *IQRs*, and plot them versus *Δt*. Figure 3 show the results for the PULI station, and the other stations provide similar results. According to this sensitivity analysis, the medians and *IQRs* of |*S*| and *K* are relatively stable when *Δt*≥1000 days. Hence, we consider *Δt*=1000 days to estimate the upper thresholds hereafter.

The parameter $A_{thr}$ controls the value of the upper thresholds. For example, consider the N- and E-components of |*S*| series (|*S*|$_N$ and |*S*|$_E$ for short) for the PULI station: we estimate the upper thresholds corresponding to different $A_{thr}$∈{1:1:5}, as shown in Fig. 4. We observe that, when $A_{thr}$=1, from 2012/12/01 to 2012/12/30, four anomalous points are found in |*S*|$_N$ and 11 anomalous points in |*S*|$_E$ prior to the 2012/12/31, $M_L$=5.28 earthquake, whose distance to the PULI station is within 60 km. However, some of those anomalies are not anomalous anymore when increasing $A_{thr}$. Similar results are observed for the 2013/03/27, $M_L$=6.24 and the 2013/06/02, $M_L$=6.48 earthquakes. Moreover, when $A_{thr}$=4, fragmentary anomalies still appear prior to the 2013/06/02, $M_L$=6.48 earthquake within 53 days, while there are no anomalies prior to the 2013/03/27, $M_L$=6.24 earthquake. We explain in a later section how to determine the optimal value of $A_{thr}$ for a given dataset.

### 3.3 Daily numbers of anomalous indices versus earthquakes

At each station, the four indices (|*S*|$_N$, |*S*|$_E$, $K_N$, $K_E$) are used to state if a day is labeled as anomalous. First, at any day, we estimate the Anomaly Index Number (AIN), which is the number of indices greater than their upper thresholds. Figure 5a shows the time series of *AIN* when $A_{thr}$=1 for the PULI station. We observe that the *AIN* prior to strong earthquakes might be larger than during periods of relatively seismic quiescence.

On the other hand, we label a day as anomalous if the *AIN* of that day is greater than or equal to a threshold number ($N_{thr}$). Figure 5b shows, for $A_{thr}=1$, the days labeled as anomalous corresponding to different values of $N_{thr} \in \{1:1:4\}$ for the PULI station. It is obvious that the number of anomalous days decreases with increasing $N_{thr}$.

## 3.4 Spatial range of geoelectric precursors

The recorded geoelectric fields at a given station are affected by instruments themselves, local human activities, geological and hydrological structures, and so on. Thus, provided that rock fracturing generates electric signals, we expect that stations could miss the fracture-induced electric signals, possibly overprinted by other transients or screened during their propagation. Moreover, the amplitudes of the fracture-induced electric signals may decay with distance, limiting again the detection potential of each station.

The critical region radius of correlated seismicity before final events is found as a function of magnitude (Bowman et al., 1998). On the other hand, pre-seismic ultra-low frequency (ULF) emissions would be detected for a given station, which satisfy the relation between magnitude (*M*) and epicentral distance (*R*): $0.025R < M - 4.5$ (Hattori, 2004). Hence, considering $M_L \in [5, 7]$ earthquakes, the correlated distance is approximately 20-200 km seismically (see Fig. 7 of Bowman et al., 1998), and 20-100 km geomagnetically. Figure 6 shows the time series of days labeled as anomalous at the PULI station for $A_{thr}=1$ and $N_{thr}=1$. $M_L \geq 5$ earthquakes are also plotted, selected when the source-to-station distance is smaller than or equal to different cut-off distances ($R_c \in \{20:10:100\}$ km). The number of selected earthquakes naturally increases with $R_c$. Comparing the selected earthquakes to the anomalous days for different $R_c$ values, we discover that, when $R_c \geq 70$ km, the number of earthquakes with few preceding anomalous days increases. For instance, when $R_c=70$km, the 2012/08/14, $M_L=5.18$ and 2012/12/3, $M_L$=5.02 earthquakes appear with no or few preceding anomalous days. When $R_c$=90 km, the 2012/8/18, $M_L$=5.2 earthquake appears with no anomalous days. The larger $R_c$ is, the more events are included.

## 3.5 Observation time windows and numbers of anomalous days versus earthquakes

The number of anomalous days prior to large earthquakes is different from event to event. Figure 7 shows the days labeled as anomalous for $A_{thr}$=1 and $N_{thr}$=1 at the PULI station. We target the period from 2013/03/01 to 2013/03/28 in detail, which includes the only 2013/03/27, $M_L$=6.24 earthquake that occurred near that station. First, we focus on the window from 2013/03/24 to 2013/03/27 with an observation time length $T_{obs}$=4 days: one anomalous day within this window is thought to correspond to the earthquake, providing the anomaly ratio: 1/4=0.25. For $T_{obs}$=10 days (from 2013/03/18 to 2013/03/27), three anomalous days seem to be related to the earthquake, hence the anomaly ratio: 3/10=0.3; for $T_{obs}$=14 days (from 2013/03/14 to 2013/03/27), anomaly ratio=6/14=0.43; for $T_{obs}$=16 days (from 2013/03/12 to 2013/03/27), anomaly ratio=7/16=0.44. When the ratio of the number of enclosed anomalous days to the length of an observation window is greater than or equal to a threshold ratio ($P_{thr}$), we consider that the anomalous days are related to the earthquake. On the other hand, we also observe that an earthquake does not occur immediately after a given anomalous period, as shown in Fig. 7, which is also confirmed by previous studies (Chen et al., 2017; Han et al., 2017). The time gap between the end of an anomalous time window and the event is referred to as a leading time window ($T_{lead}$ in day).

In summary, we consider six main parameters ($A_{thr}$, $N_{thr}$, $R_c$, $T_{obs}$, $T_{thr}$, $T_{lead}$) to study the relationships between geoelectric anomalies and earthquakes. The threshold ratio $P_{thr} = \frac{T_{thr}}{T_{obs}}$ is also taken into account. We further build up a predictive model based on these parameters and optimize the model parameters (see also Fig.3 of Chen et al., 2017).

# 4. Examination of the relationship between anomalies and earthquakes

Building on the GEMS network and the concept of 'Time of Increased Probability (TIP),' a GEMSTIP algorithm has been defined (see Chen et al., 2017; Chen and Chen, 2016), which is used to identify and test relationships between geoelectric statistical anomalies and earthquake occurrences. In this study, we improve the GEMSTIP algorithm from a single station method to a joint stations method, propose another evaluation of model performance, and define a confidence bound for significance tests. This new version of the GEMSTIP algorithm consists of supervised machine learning and binary classification, meaning that the algorithm learns an optimal model from a training dataset, and predicts future events as one adds new data. The training dataset consists of known input data (geoelectric statistical anomalies) and corresponding output data (earthquakes). For this supervised learning, the machine extracts the features of geoelectric anomalies to predict earthquakes. This algorithm comprises two main parts. In the first one, we establish a predictive model. This model labels days as anomalous based on given anomalous statistical indices, and then provides 'TIP' alarms based on those anomalies. In the second part, we evaluate the model by comparing TIP alarms with observed earthquakes temporally and spatially. Before data analysis, we divide the dataset into two independent subsets: the training set and the validation set. The algorithm evaluates the fit of predictive models using different model parameters on the training dataset, and selects the optimal parameters. The models with the optimal parameters are then used to process the validation dataset, and provide the forecasting scores. Conducting significance tests, we assess the practicability of the predictive model for the forecasting strategy.

## 4.1 GEMSTIP algorithm: Single station method
### 4.1.1 Establishing a predictive model

In Section 3, we presented the factors used to define earthquake alarms based on geoelectric anomalies. For the sake of generalization, we take into account the magnitude of an event ($M_c$) and the length of a predictive window ($T_{pred}$). Hence, the model parameter vector $\boldsymbol{g}$ of each station is as follows:

$$\boldsymbol{g} = [M_c, R_c, A_{thr}, N_{thr}, T_{thr}, T_{obs}, T_{lead}, T_{pred}]. \tag{3}$$

This predictive model possesses eight parameters, and the illustrative meaning of each parameter is shown in Fig. 8. $M_c$ is the minimum event magnitude that can be predicted. It is convenient for calculation to introduce $T_{thr} = \lceil P_{thr} * T_{obs} \rceil$, where $P_{thr}$ is a threshold ratio ranging from 0.1 to 0.5, and $\lceil x \rceil$ means the ceiling of $x$, i.e. the smallest integer greater than or equal to $x$.

Considering a parameter vector $\boldsymbol{g}$, an index of target earthquakes is define:

$$Q(t|\boldsymbol{g}) =$$
$$\begin{cases} Q(t = t_i) = 1, \\ \quad \text{if } M_{Li} \geq M_c \cap \|(x_i, y_i, z_i) - (x_{sta}, y_{sta}, 0)\| \leq R_c, i = 1 \text{ to } N_{EQ}, \\ Q = 0, \quad \text{otherwise} \end{cases} \tag{4}$$

where $Q$ is the time series of 0 (no target earthquake) or 1 (target earthquake), $t$ is the time in days; $(x_i, y_i, z_i)$, $t_i$, and $M_{Li}$ are the location, occurrence time (in days), and magnitude of the $i^{th}$ target earthquake, respectively; $N_{EQ}$ is the number of earthquakes, and $(x_{sta}, y_{sta})$ is the location of the considered station. $M_c$ is the cut-off magnitude of target earthquakes, and $R_c$ is the cut-off distance of target earthquakes to a station as we select earthquakes within a source-to-station distance smaller than or equal to $R_c$, as shown in Fig. 8a. In order to obtain TIPs, we begin with the 'Anomaly Index Number,' which is defined as:

$$F_{AIN}(t|\boldsymbol{g}) = \sum_{i=1}^{2}\left(I\left(|S_i(t)| > \theta(t;|S_i|)\right) + I\left(K_i(t) > \theta(t;K_i)\right)\right), \quad (5)$$

where $F_{AIN}$ can take up to five values (from 0 to 4), $t$ is the time in days, $i$ is the N- or E-component, $\theta$ is the threshold value defined by Eq. (2), and $|S|$ and $K$ are the absolute value of skewness and kurtosis, respectively, and $I(\Omega)$ is a logical function:

$$I(\Omega) = \begin{cases} 1, & \Omega \text{ is true} \\ 0, & \text{otherwise} \end{cases}. \quad (6)$$

The 'Anomalous Time' is then defined as:

$$F_{AT}(t|\boldsymbol{g}) = I(F_{AIN}(t|\boldsymbol{g}) \geq N_{thr}), \quad (7)$$

where $F_{AT}$ is the time series of 0 (non-anomalous time) or 1 (anomalous time), $t$ is the time in days, and $N_{thr}$ is an integer threshold number smaller than or equal to 4. The 'Sum of Anomalous Time' within a moving observation time window ($T_{obs}$) can be defined as:

$$F_{SAT}(t|\boldsymbol{g}) = \sum_{i=t-T_{obs}+1}^{t} F_{AT}(t_i|\boldsymbol{g}), \quad (8)$$

where $F_{SAT}$ is an integer time series, $t$ and $t_i$ are times in days, $T_{obs}$ (in days) is the length of the observation time window. The 'Time of Increased Probability (TIP),' illustrated in Fig. 8b, is thus defined as:

$$T_{TIP}(t|\boldsymbol{g}) = \begin{cases} T_{TIP}(t = t_i + T_{lead} \text{ to } t_i + T_{lead} + T_{pred}) = 1, & \text{if } F_{SAT}(t_i) \geq T_{thr} \\ T_{TIP}(t = t_i + T_{lead} \text{ to } t_i + T_{lead} + T_{pred}) = 0, & \text{if } F_{SAT}(t_i) < T_{thr} \end{cases}, \quad (9)$$

where $T_{TIP}$ is the time series of 0 (non-TIP) or 1 (TIP), $t$ and $t_i$ are the times in days, and $T_{thr}$ (in days) is a threshold number. We issue alarms between $t_i+T_{lead}$ and $t_i+T_{lead}+T_{pred}$ when $F_{SAT} \geq T_{thr}$ at time $t_i$. The value ranges of the model parameters are listed in Table 2. We adopt the concept of a grid search, meaning that hundreds of thousands of parameter vectors are generated within the value ranges. We then evaluate the score on a training dataset for each vector considered as a fit to the observations.

## 4.1.2 Scoring of models

The score of each model parameter vector is evaluated as:

$$d(\boldsymbol{g}) = 1 - \tau(\boldsymbol{g}) - n(\boldsymbol{g}). \tag{10}$$

The function $\tau(\boldsymbol{g})$ is:

$$\tau(\boldsymbol{g}) = \frac{\sum_t I(T_{TIP}(t|\boldsymbol{g})=1)}{\sum_t I(T_{TIP}(t|\boldsymbol{g})\geq 0)}, \tag{11}$$

which is the fraction of alarmed time cells. The function $n(\boldsymbol{g})$ is:

$$n(\boldsymbol{g}) = \frac{\sum_t I(T_{TIP}(t|\boldsymbol{g})=0 \cap Q(t|\boldsymbol{g})=1)}{\sum_t I(T_{TIP}(t|\boldsymbol{g})\geq 0 \cap Q(t|\boldsymbol{g})=1)}, \tag{12}$$

which is the fraction of missed earthquakes. The smaller $\tau(\boldsymbol{g})$ and $n(\boldsymbol{g})$, the better, meaning that the model catches target earthquakes with a high success rate and a precise time resolution. The two functions are similar to Molchan scores (see Molchan, 1997), for which we can also provide confidence intervals as proposed by Zechar and Jordan (2008). Figure 9 shows the scatterplot of ($\tau$, $n$) for the PULI station during the training phase from its onset to 2015/03/31. Each dot corresponds to the performance of a single model as defined above.

For comparison, Chen and Chen (2016) and Chen et al. (2017) use C1 and F1 scores, measuring the ratios of true negatives and true positives, respectively, to the sum of false negatives, false positives and themselves. They introduced one more time parameter for coarse-graining earthquake occurrence times in order to increase discrimination on the F1 axis (see Chen & Chen, 2016). In this study, the usage of Molchan scores avoids introducing this redundant time parameter, which is generated only for mathematical purposes rather than from physical observations.

The fit between the two time series of $Q$ and $T_{TIP}$ is quantified by $d \in [-1,1]$. The case $d>0$ means that the model successfully and non-randomly forecasts events, whereas $d \leq 0$ means that the prediction of the model is no better than a random guess (=0) or even worse than random. We thus rank the model parameter vectors according to their corresponding $d$ values, select the top 10 model parameter vectors of each station, and further analyze their performances on validation datasets.

## 4.2 GEMSTIP algorithm: Joint stations method
### 4.2.1 Description

In the joint stations method, the selected optimal parameter vectors of each station (see previous section) will be combined in order to build a parameters tensor for the Taiwan area. In this way, the spatial variations of $Q$ and $T_{TIP}$ are also considered. A model with a given parameters tensor is then scored using the training and validation datasets in order to evaluate the overall fit for the Taiwan area. Figure 10 shows the schematic diagram describing the joint stations method. For example, station B in Fig. 10 misses event E2, so that event E2 is not predicted in the single station method applied to that station. However, station C hits event E2 successfully, so that event E2 is predicted when considering the joint stations method of both stations B and C. A similar case is shown for event E4 and stations A and C. The advantage of the joint stations method is that, if an earthquake occurs within the detectable distances ($R_c$) of two stations and that one station issues an alarm for the earthquake, while the other does not, the joint stations method still considers the alarm to be successful. This non-simultaneous alarms of stations might result from electric preferential propagations due to rupture directivity (Ogawa et al., 1985), migration of seismic activity (Sanders, 1993; Wu et al., 2008, 2011), and conductive structure complexity (Bertrand et al., 2009; Bertrand et al., 2012; Huang, 2011). Hence, the results of the joint stations method should be relatively more stable compared with the single station method.

The selected optimal vector $g$ of each station composes the parameter tensor $G$ for the Taiwan area, which is described as follows:

$$G = \{g_i, \quad i = 1 \text{ to } N_{sta}\}, \tag{13}$$

where $N_{sta}=20$ is the number of all stations in this paper.

The formal description of the joint stations method is similar to that of the single station method. The expressions of $Q$ and $T_{TIP}$ for the joint stations method include spatial variables, use the data of all stations, and are modified as follows:

$$Q(x, y, t|G) = \begin{cases} Q(x = x_{i,j}, y = y_{i,j}, t = t_{i,j}) = 1, \\ \text{if } M_{Li,j} \geq M_{cj} \cap \|(x_{i,j}, y_{i,j}, z_{i,j}) - (x_j, y_j, 0)\| \leq R_{cj}, i = 1 \text{ to } N_{EQj}, j = 1 \text{ to } N_{sta} \\ Q = 0, otherwise \end{cases} \tag{14}$$

In Eq. (14), $Q$ is the space and time grids of 0 (no target earthquakes, or no pentagrams in Fig. 10) or 1 (target earthquakes, or pentagrams in Fig. 10); $(x_{i,j}, y_{i,j}, z_{i,j})$, $t_{i,j}$, and $M_{Li,j}$ are the location, occurrence (in days), and magnitude of the $i^{th}$ earthquake for the $j^{th}$ station, respectively; $(x_j, y_j)$ is the location of the $j^{th}$ station; $N_{EQj}$ is the number of selected earthquakes for the $j^{th}$ station, as we select earthquakes with magnitude greater than or equal to a cut-off magnitude for the $j^{th}$ station ($M_{cj}$) within a source-to-station distance smaller than or equal to a cut-off distance for the $j^{th}$ station ($R_{cj}$).

$$T_{TIP}(x, y, t|G) = \begin{cases} T_{TIP}(\|(x, y) - (x_j, y_j)\| \leq R_{cj}, t = t_i + T_{leadj} \text{ to } t_i + T_{leadj} + T_{predj}) = 1, \\ \text{if } F_{SATj}(t_i) \geq T_{thrj}, j = 1 \text{ to } N_{sta} \\ T_{TIP}(\|(x, y) - (x_j, y_j)\| \leq R_{cj}, t = t_i + T_{leadj} \text{ to } t_i + T_{leadj} + T_{predj}) = 0, \\ \text{if } F_{SATj}(t_i) < T_{thrj}, j = 1 \text{ to } N_{sta} \\ T_{TIP} = nan, \quad otherwise \end{cases} \tag{15}$$

In Eq. (15), $T_{TIP}$ is the space and time grids of 0 (non-TIP, or white regions within a dotted circle of a station in Fig. 10), 1 (TIP, or yellow regions in Fig. 10), or not-a-number (nan, or white regions out of dotted circles in Fig. 10). For the $j^{th}$ station, we issue alarms within a distance $R_{cj}$ km between $t_i+T_{leadj}$ and $t_i+T_{leadj}+T_{predj}$ when $F_{SATj} \geq T_{thrj}$ at time $t_i$. On the other hand, the evaluation of the agreement between $Q$ and $T_{TIP}$ for the joint stations method are similar to Section 4.1.2, which is conditioned to $\boldsymbol{G}$ instead of $\boldsymbol{g}$:

$$\tau(\boldsymbol{G}) = \frac{\sum_x \sum_y \sum_t I(T_{TIP}(x,y,t|\boldsymbol{G})=1)}{\sum_x \sum_y \sum_t I(T_{TIP}(x,y,t|\boldsymbol{G})\geq 0)},$$

$$n(\boldsymbol{G}) = \frac{\sum_x \sum_y \sum_t I(T_{TIP}(x,y,t|\boldsymbol{G})=0 \cap Q(x,y,t|\boldsymbol{G})=1)}{\sum_x \sum_y \sum_t I(T_{TIP}(x,y,t|\boldsymbol{G})\geq 0 \cap Q(x,y,t|\boldsymbol{G})=1)},$$

$$D(\boldsymbol{G}) = 1 - \tau(\boldsymbol{G}) - n(\boldsymbol{G}). \tag{16}$$

Note that the score $D$ is used for the joint stations method, while $d$ stands for the single station method.

### 4.2.2 Significance: Confidence bound

Zechar and Jordan (2008) developed a confidence bound for Molchan error diagrams, which is derived from a null hypothesis and a binomial distribution, as follows:

$$\sum_{n=h}^{N} B(n|N, \widetilde{P_A}) \leq \alpha, \tag{17}$$

where $B$ is a binomial distribution, $h$ is the number of hit events, $N$ is the total number of events, $\widetilde{P_A}$ is a reference probability of an alarmed region, $\alpha$ is a significance level. Equation (17) predicts that the alarm-based prediction has a significant skill if the probability of obtaining $h$ or more hits by chance is less than or equal to $\alpha$.

By solving Eq. (17), we obtain not only a confidence bound of ($\tau_{CB}$, $n_{CB}$), but also $D_{CB}$ by the relation 1-$\tau_{CB}$-$n_{CB}$. We then define the maximum of the boundary value $D_{CB}$:

$$D_{CB}^{max}(N_{EQ}, \alpha) = max\{D_{CB}(N_{EQ}, \alpha)\}, \tag{18}$$

where $N_{EQ}$ is the number of total earthquakes, and a significance level $\alpha$=0.05 is set in this paper. Equation (18) thus predicts that a predictive model is practical if its performance $D$ is larger than $D_{CB}^{max}$. Figure 9 shows an example of the confidence bound of ($\tau_{CB}$, $n_{CB}$) for $N_{EQ}$=10 and $\alpha$=0.05; see Fig. 3 of Zechar and Jordan (2008) for more confidence bounds for different values of $N_{EQ}$ and $α$.

## 4.3 Results

According to the sensitivity analysis of Section 3.1, we opt for $Δt$=1000 days to define the threshold $θ(t; y)$ of a given index $y$. We define a training phase from the onset time of each station up to 2014/06/30, labelled as Trn of case 01 in Table 3 (see below for the definition of all cases). For that period, fifteen stations have recorded more than 1000 days of data. In order to study the effect of the lengths of training phases, the training phases are also extended by three months case by case, as shown in Table 3; that is, the ending times of the training phases are 2014/06/30, 2014/09/30, 2014/12/31, 2015/03/31, 2015/06/30, 2015/09/30, and 2015/12/31, which amount to seven cases (labelled from 01 to 07 in Table 3). On the other hand, the lengths of validation phases are tested by selecting 3, 6, 9, and 12 months following their corresponding training phases (labelled as Vld03, Vld06, Vld09, and Vld12, respectively). The number of earthquakes in each phase is also listed as the figure in the bracket of Time column in Table 3.

First of all, using a single station method described in Section 4.1, we estimate the $d$ values of different parameter vectors $g$ for each station within the training phase of case 01. According to the $d$ values for each station, we select the top ranking parameter vectors. In this study, we select for each station the top 10 outstanding vectors. Subsequently, we combine the top 1 vectors of all stations into the first parameter tensor $G_1$ for the Taiwan area, the top 2 vectors of all stations into the second parameter tensor $G_2$, and so on, which amounts to 10 sets of parameter tensors $G_i$ for $i$=1 to 10. Using the joint stations method described in Section 4.2, we then estimate the overall $D$ values for the Taiwan area within the training phase and its following four kinds of validation phases for case 01. There are 10 $D$ values for each phase, and they are reported as mean

± 2 standard deviations in Fig. 11. Table 3 also lists the statistics of ($\tau$, $n$, $D$). Repeating the abovementioned procedure, we obtain the results for cases 02-07. We find out that the average values of $D$ for the training phases of cases 01-06 are similar, approximately 0.85; however, in case 07, it decreases to 0.79. For the standard deviations of $D$ in the training phases, they are relatively small in cases 04-06, approximately 0.015, while the others are on average 0.088. In the validation phases for all cases, we observe a trend in the $D$ values, i.e. Vld03>Vld06>Vld09>Vld12, meaning that the forecasting time period cannot be too long, as its score decreases with its duration. The mean of the $D$ values in the cases 04 and 05 of Vld03 and the case 04 of Vld06 (average 0.885) are the largest, and their standard deviations (average 0.017) are the smallest. The cases 04 and 05 perform well both in the training and validation phases. The mean ± standard deviations of the $D$ values for cases 04 and 05 in the training and validation phases are similarly comparable, meaning that the predictive model could be made operational.

As shown above, the fitting scores in cases 04 and 05 perform well in both the training and validation phases, suggesting that the best length of the training phase for optimizing the parameter vectors is approximately between 1000 and 1200 days. More rigorous statistical tests will be implemented when collecting more data. In this way, we would divide the datasets into several non-overlapping segments according to different lengths of the training phase, estimate $D$ values for each segment, and then analyze their statistics. Except for the cases 01, 05, 06, and 07 of Vld12, the ranges of $D$ for both the training and validation datasets are above the confidence bound $D_{CB}^{max} = 0.46$ for $N_{EQ}$=5 and $\alpha$=0.05. Note that the number of target earthquakes of most phases is larger than 5, so that the $D_{CB}^{max}$ value for more earthquakes would be much lower. This means that the predictive model proposed in this study is meaningful and could be put in practice, i.e. there is a significant 'hidden correlation' in the seismoelectric pattern.

# 5. Frequency bands of earthquake-related geoelectric signals

As noise is ubiquitous in Nature, it is an important issue to determine frequency bands of earthquake-related signals. Increasing the earthquake-related signal-to-noise (S/N) ratio improves the study of earthquake precursors, and promotes the accuracy of precursor-based earthquake probability forecasts. Filtering time series is a way to increase the S/N ratio. In previous studies, electromagnetic signals are analyzed in the 0.001-0.01 Hz frequency range, without justification (see Han et al., 2017; Hattori et al., 2013; Uyeda et al., 2002). We propose here to study the influence of the underlying frequency bands on the performance of the forecasting scheme.

## 5.1 Filtering

An ideal filter belongs to one of the three main types as follows:

1. Ideal low-pass filter:

$$|H(f)| = \begin{cases} 1, |f| < f_c \\ 0, |f| > f_c \end{cases}, \tag{19}$$

where $|H(f)|$ is the gain in the frequency domain, and $f_c$ is a cut-off frequency in Hz.

2. Ideal high-pass filter:

$$|H(f)| = \begin{cases} 0, |f| < f_c \\ 1, |f| > f_c \end{cases}. \tag{20}$$

3. Ideal band-pass filter:

$$|H(f)| = \begin{cases} 1, f_1 < |f| < f_2 \\ 0, otherwise \end{cases}, \tag{21}$$

where $f_1$ and $f_2$ are relatively low and high cut-off frequencies in Hz, respectively.

In this study, we use Butterworth filters (see Butterworth, 1930). Their main properties are that the frequency response is maximally flat in the passband, and gradually rolls off toward zero in the stopband. Its response is flat, close to DC signals, decays to -3dB at the cut-off frequency, and decreases with a decaying rate of - 20$n$dB/decade, where $n$ is the number of poles in the filter. The Butterworth filter is suitable for analyzing low-frequency signals. In this study, we use three-order low-, high-, and band-pass Butterworth filters.

## 5.2 Results of the GEMSTIP analysis

In this study, the sampling rate of geoelectric fields analyzed is 1 Hz. Hence, the Butterworth filters are applied with different cut-off frequencies from $10^{-4}$ to $10^{-0.5}$ Hz with a step of 0.25 in log scale. When loading the raw data of the geoelectric fields, we first apply the low- and high-pass filters with different cut-off frequencies. Next, we calculate the skewness and kurtosis of the filtered data, and repeat the analysis of the GEMSTIP algorithm proposed in Section 4. Using the GESMTIP results for the low- and high-pass filtered datasets, we can thus determine the optimal frequency bands of the earthquake-related signals with high S/N ratio.

### 5.2.1 Results with low- and high-pass filtered data

Based on Section 4.3, we select the optimal training phase from the onset time of each station up to 2015/03/31, and the optimal validation phase from 2015/04/01 to 2015/06/30. To begin with the low-pass filtered datasets, using the GESMTIP algorithm, we first get the mean and standard deviations of the $D$ values versus the different cut-off frequencies, as shown in Fig. 12 (red lines). The mean ± 2 standard deviations of the $D$ values for both the training and validation phases are very close, suggesting again that the algorithm is robust and that its optimal parameters tensors are informative. On the other hand, the mean of the $D$ values remains relatively stable and high (approximately 0.85) with $f_c \leq 10^{-1.75}$ Hz (Period $T \geq 56$ sec). The mean shows a slightly decreasing trend from ~0.85 to ~0.78 when $f_c > 10^{-1.75}$ Hz, and the standard deviations become much larger. This suggests that the earthquake-related signals would be contaminated by noise at higher frequencies.

Subsequently, the above mentioned procedure is carried out on the high-pass filtered datasets. The $D$ values versus different cut-off frequencies are also shown in Fig. 12 (blue lines). Only for the high-pass filtered dataset with $f_c=10^{-4}$ Hz are the mean ± 2 standard deviations of the $D$ values comparable in the training and validation phases. The gaps between the mean of the $D$ values for the training and validation phases for the high-pass filtered datasets with $f_c \geq 10^{-3.75}$ Hz are larger than those for the low-pass filtered datasets. Furthermore, the standard deviations for the high-pass filtered datasets with $f_c \geq 10^{-3.75}$ Hz are much larger than those for the low-pass filtered datasets. The results suggest that the ultra-low frequency ($f<10^{-3.75}$ Hz or $T>5623$ sec) signals stabilize the performance of the prediction model, also confirming that the earthquake-related signals are seriously disturbed by high frequency noise.

Earthquake-related signals might exist at all frequencies because the $D$ values for the low- and high-pass filtered datasets for both the training and validation phases are all larger than the confidence bound ($D_{CB}^{max} = 0.46$ for $N_{EQ}$=5 and $\alpha$=0.05). However, the high frequency noise has more power than the earthquake-related signals. Hence, the $D$ values for the signals including higher frequency components perform worst. It thus seems necessary to first low-pass the original signal before processing it with the GEMSTIP methodology.

### 5.2.2 Results with optimal band-pass filtered data

Following Section 5.2.1, we can determine one optimal frequency band from $f_1=10^{-4.0}$ to $f_2=10^{-1.75}$ Hz ($T=\sim56$ sec – $\sim2.78$ hr), and the band-pass filtered data in this band is called Bapass4.0. We select one additional frequency band from $f_1=10^{-3.5}$ to $f_2=10^{-1.75}$ Hz ($T=\sim56$ sec – $\sim52.7$ min) as a control group, and the data in this band is called Bapass3.5. Then, we calculate the skewness and kurtosis of the Bapass3.5 and Bapass4.0 datasets. We select again the same training phase (denoted as Trn) as in Section 5.2.1, and the four validation phases following the training phase (denoted as Vld03, Vld06, Vld09, and Vld12). Through the GEMSTIP analysis procedure, we first obtain the optimal models of each station. The top 10 model parameters of all stations are categorized in Tables S1, S2, and S3 of the Supplementary Materials for the raw, Bapass3.5, and Bapass4.0 datasets, respectively. Figure 13 shows the probability density functions (PDFs) of the top 10 parameters for all stations. We find that the optimal distance ($R_c$) of a signal to a station on average 50-60 km, the anomalous number threshold ($N_{thr}$) is in the range 1-4, the index value threshold ($A_{thr}$) is 1-2, the anomalous day threshold ($T_{thr}$) is in the range 1-5 days, the observation window ($T_{obs}$) is smaller than 20 days, the ratio threshold ($P_{thr} = \frac{T_{thr}}{T_{obs}}$) is 0.1-0.2, and the leading time window ($T_{lead}$) is non-zero. Such non-zero $T_{lead}$ values indicate pre-seismic electromagnetic quiescence, which might be caused by the variation of the constitutive electrokinetic parameters, such as underground resistance, capacitance, and inductance (see our other article presenting the coupled mechano-geoelectric COS model in this special volume).

We further get the mean and standard deviations of the $D$ values of the Bapass3.5 and Bapass4.0 datasets for the five different phases, as shown in Fig. 14. Table 4 also lists the statistics of ($\tau$, $n$, $D$) for the two datasets. The $D$ values for the raw data for the five phases are also shown in Fig. 14 as reference, and have been obtained in Section 4.3. The standard deviations of the $D$ values of the different phases for the Bapass4.0 data are all smaller than those for the raw and Bapass3.5 data, especially for the Vld12 phase, suggesting that the earthquake-related signals with $f \lesssim 10^{-3.5}$ Hz are more significant and more informative than others, an important information to constrain the predictive model. The mean ± 2 standard deviations of the $D$ values for the Bapass4.0 data are almost the same for the Trn, Vld03, Vld06, Vld09, and Vld12 phases, suggesting that the optimal parameters tensors obtaining from the training datasets including low frequency components and moderately high frequency components could be used to forecast for a longer future period. In summary, the GEMSTIP algorithm using the two band-pass filtered datasets is more robust compared with using the raw data. Especially for the geoelectric signals with the optimal frequency band $10^{-4.0} \leq f \leq 10^{-1.75}$ Hz, they are more strongly correlated to earthquakes.

### 5.3 Precursor-based earthquake probability forecasts

Based on the above, earthquake probability forecasts depended on pre-seismic geoelectric anomalies can be constructed. This earthquake probability is an ensemble of probabilities conditioned to top ranking models of all stations, which are the products of hit rates and binary numbers {0, 1} of predicted spatio-temporal areas. In order to build up a precursor-based probability of a future event, for a given parameters tensor $G$, the hit rate $v$ is defined as:

$$v(G) = 1 - n(G), \tag{22}$$

where $n$ is the missed rate of earthquakes defined in Eq. (12). Then, the earthquake probability $P(x, y, t)$ is defined as:

$$P(x, y, t) = \frac{1}{N_{top}} \sum_{i=1}^{N_{top}} v(G_i) \cdot T_{TIP}(x, y, t | G_i), \tag{23}$$

where $N_{top}=10$ is the number of the top model parameters tensors used, $G_i$ is the $i^{th}$ optimal parameters tensors for the joint stations method, and $T_{TIP}$ is an index of 'Time of Increased Probability' described in Eq. (15). Using Eq. (23), we can estimate a probability forecast at time $t$ using the geoelectric data before time $t$. Figure 15 shows the spatio-temporal probability maps for the optimal parameters tensors obtaining from the raw, Bapass3.5, and Bapass4.0 datasets using the training phase extending from the onset time of each station up to 2015/03/31. Two target earthquakes in 2013/06/02 are located in the middle part and southern part of Taiwan, respectively, which coincide with high probabilities for the three datasets. The probabilities in the three figures increase from 2013/05/15 to 2013/06/01 before the two earthquakes, and decrease from 2013/06/03 to 2013/06/13. This illustrates a possible way to elaborate quantitative earthquake forecasts.

## 6. Conclusions

Increasing the reliability of precursor-based studies is the key to increase our ability of earthquake forecasting. The improved GEMSTIP algorithm presented in this paper is useful to test the correlations between geoelectric anomalies and earthquake occurrences. Further, using the improved GEMSTIP algorithm, we show that the frequency band $10^{-4.0} \leq f \leq 10^{-1.75}$ Hz ($T=$~56 sec – ~2.78 hr) is less contaminated by non-earthquake-related signals. Because the performing scores in the training and validation phases are close to each other and both larger than the maximal confidence bound, we conclude that there is a strong connection between geoelectric anomalies and earthquake occurrences. From this study, one might hence understand that machine learning with both geophysical anomalies and earthquakes is one possible route to improve and realize earthquake forecasting. This study lays the foundation of earthquake forecasting.

For future works, the improved GEMSTIP algorithm might be helpful to clarify the value of precursory indices obtained for specific case studies in previous works, such as detrended fluctuation analysis (Varotsos et al., 2002), natural time analysis (Varotsos et al., 2003, 2006), and principal component analysis (Uyeda et al., 2002). Another issue is the forecasting ability of geoelectric data during night-time and day-time periods, a topic our future work will focus on. In addition, the optimal length of the training phase could be determined by collecting a longer data history. Beside the analysis of field observation data, it is also necessary to improve our understanding of the coupling of both mechanics and electromagnetics using rock fracturing tests in the lab, as well as by putting efforts into theoretical and numerical physics-based modeling. We are thus also working on a conceptual physics-based model, also published in this special volume. Both approaches show encouraging results for possible seismo-electromagnetic precursor-based earthquake forecasting.

# Acknowledgements

HJC would thank Yavor Kamer for providing some insights on Molchan error diagrams. HJC and CCC are supported by Grant No. MOTC-CWB-107-E-01 from Taiwan Central Weather Bureau and by Grant No. MOST-106-2116-M-008-002 from Taiwan Ministry of Science and Technology. The authors would like to thank the anonymous reviewers and the editor for their valuable comments and suggestions.

**Figures and figure captions**

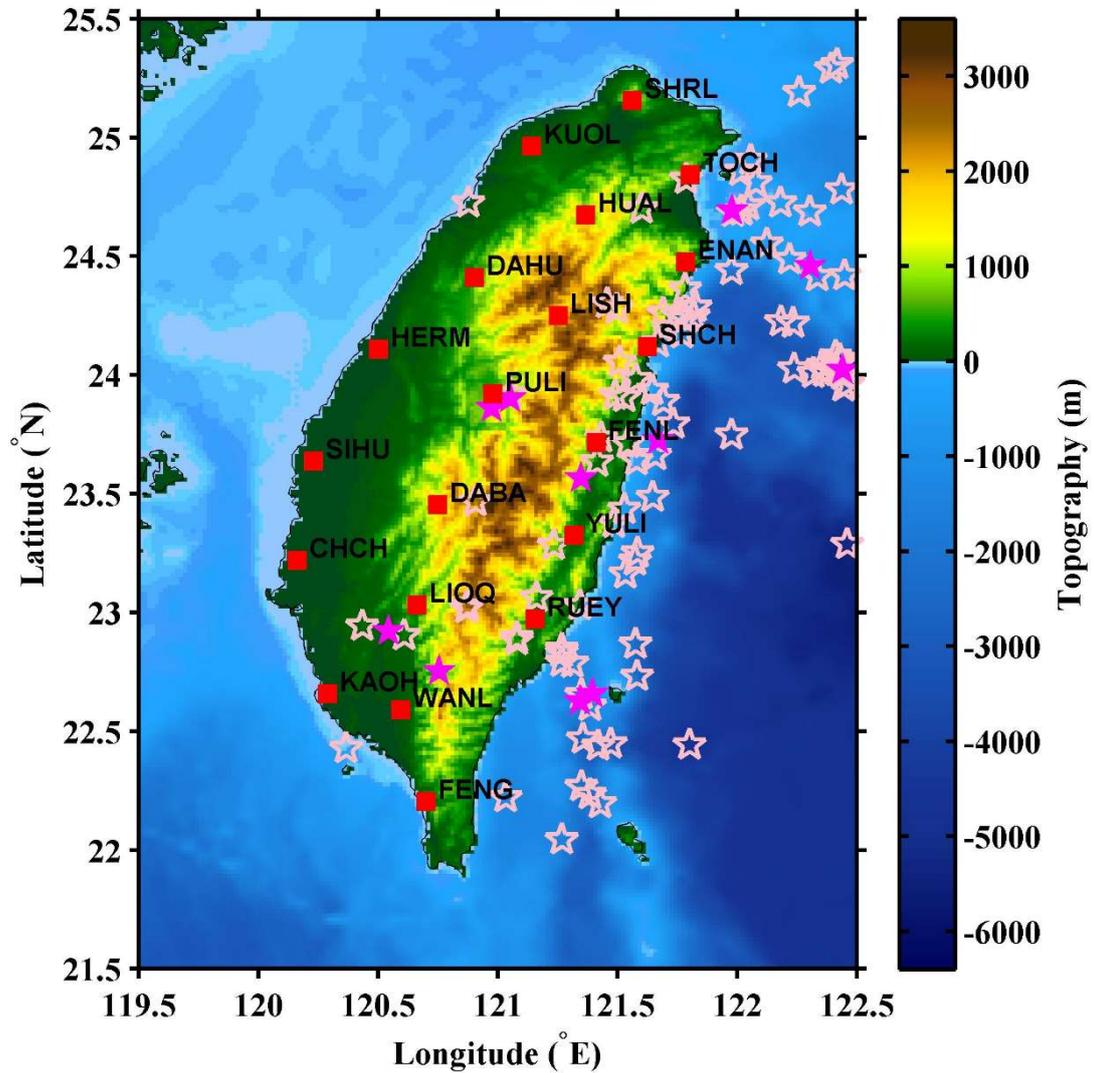

Figure 1. Spatial distributions of the geoelectric stations (red squares) and the $M_L \geq 5$ earthquakes (pentagrams). The open and filled pentagrams are $M_L \in [5,6)$ and $M_L \geq 6$ events, respectively.

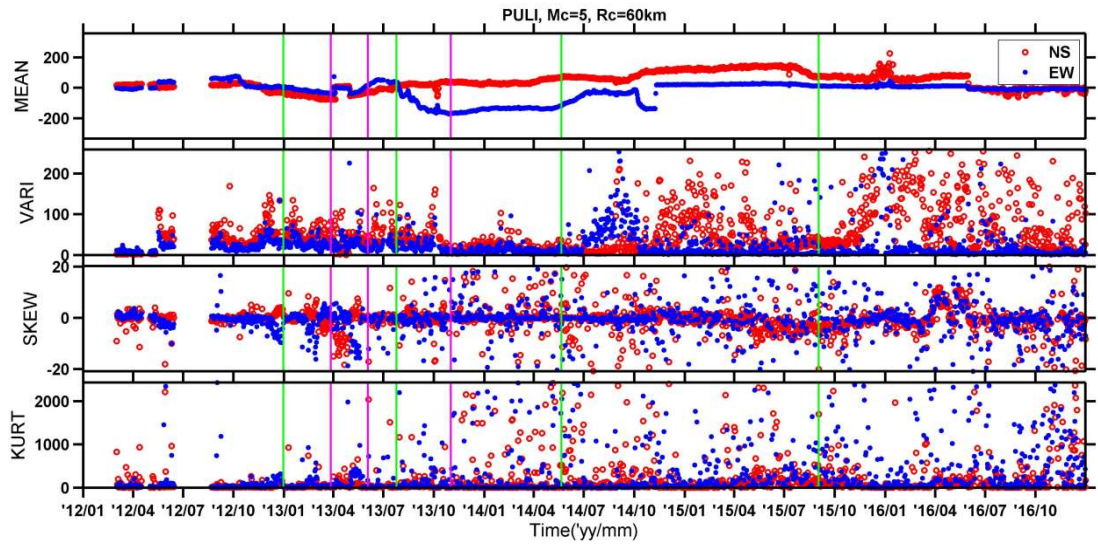

Figure 2. Time series of statistical indices and earthquakes at the PULI station. The red circles and blue dots stand for the N- and E-components, respectively. The green and magenta vertical lines indicate $M_L \in [5,6)$ and $M_L \geq 6$ events, respectively. The distances of those events to the PULI station are smaller than or equal to 60 km.

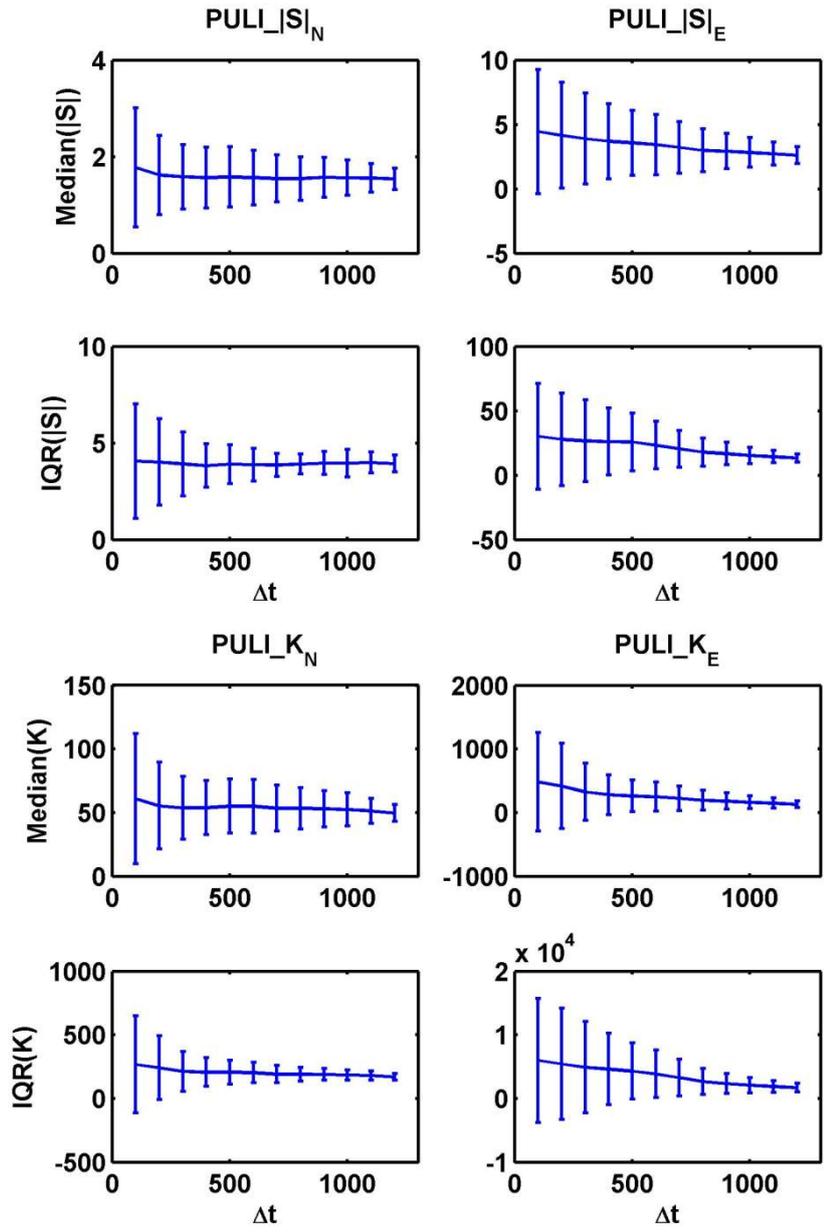

Figure 3. Sensitivity of the medians and interquartile ranges of |*S*| and *K* on *Δt* at the PULI station.

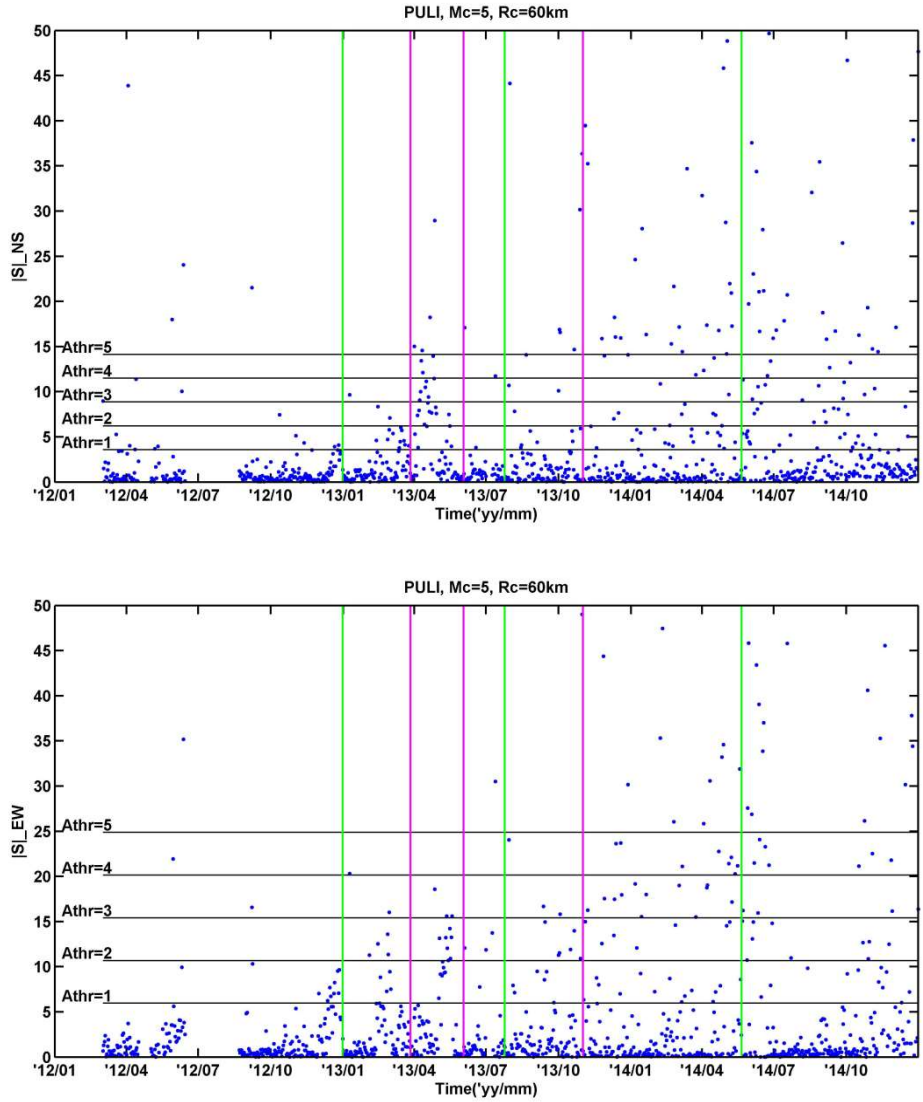

Figure 4. Time series of the absolute values of the skewness (blue dots) for the N-component (upper plot) and E-components (lower plot) for the PULI station and their upper thresholds corresponding to different $A_{thr}$ (black horizontal lines). The green and magenta vertical lines indicate $M_L \in [5,6)$ and $M_L \geq 6$ events, respectively. The distances from those events to the PULI station are smaller than or equal to 60km. Note that one time tick stands for three months.

(a)

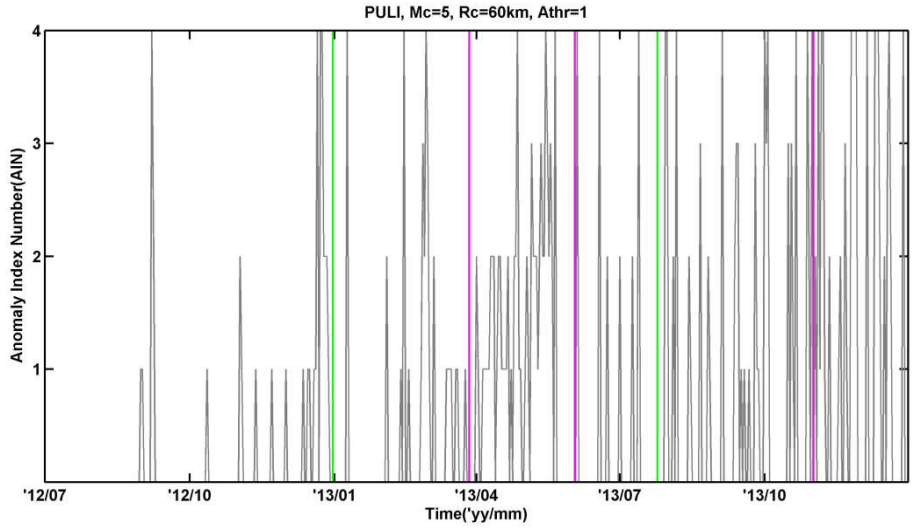

(b)

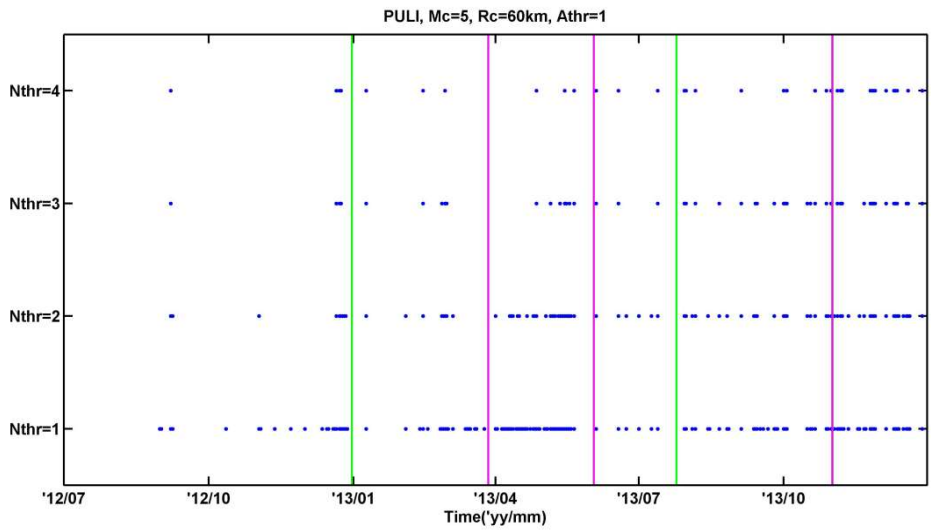

Figure 5. Time series of (a) anomaly index number (gray lines) and (b) anomaly days (blue dots) corresponding to different anomaly index threshold numbers ($N_{thr}$) using $A_{thr}=1$ at the PULI station. The green and magenta vertical lines correspond to $M_L \in [5,6)$ and $M_L \geq 6$ events, respectively. The distances of those events to the PULI station are smaller than or equal to 60 km.

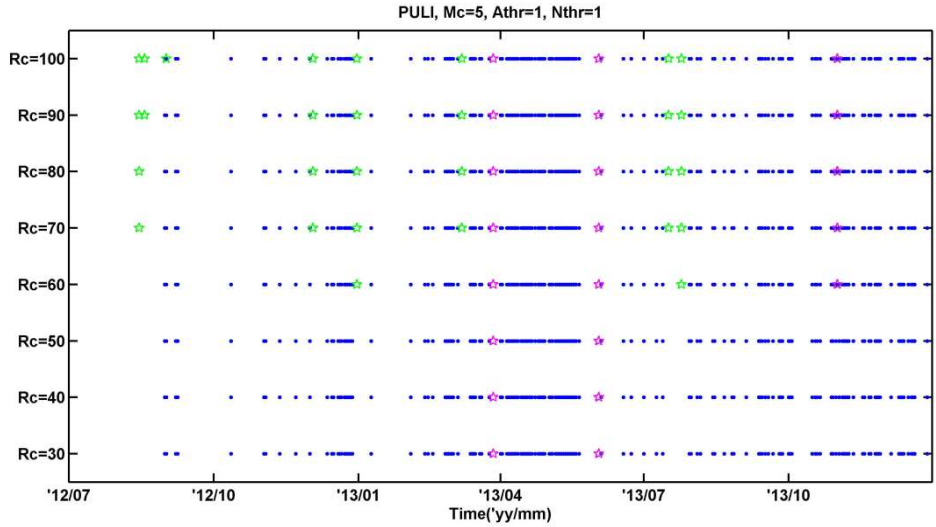

Figure 6. Time series of anomaly days together with earthquakes for different detection distances ($R_c$) at the PULI station. Anomaly days (blue dots) are defined at $A_{thr}$=1 and $N_{thr}$=1. The green and magenta stars correspond to $M_L \in [5,6)$ and $M_L \geq 6$ events, respectively. The distances of those events to the PULI station are smaller than or equal to $R_c$ km.

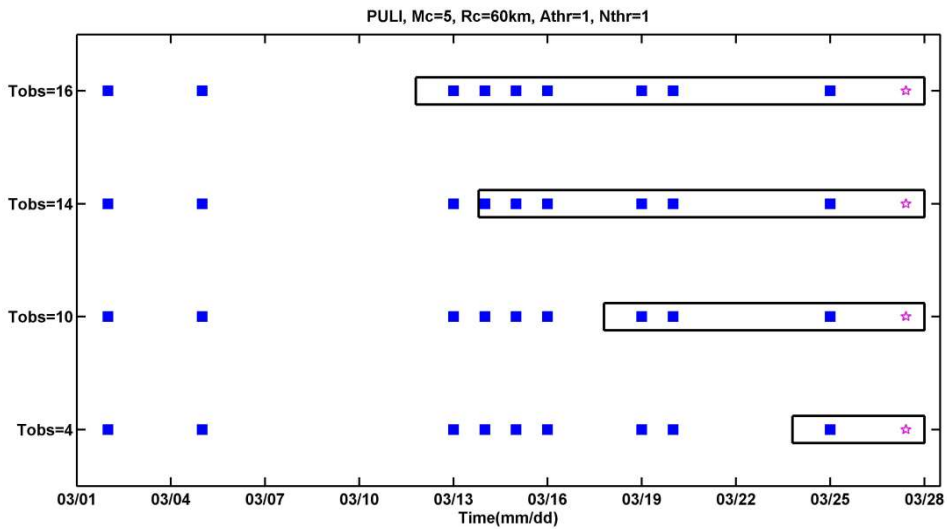

Figure 7. Time series of the lengths of the observation time window ($T_{obs}$, black rectangles) and number of anomaly days (blue squares), for the 2013/03/27, $M_L$6.24 earthquake (magenta pentagram).

(a)

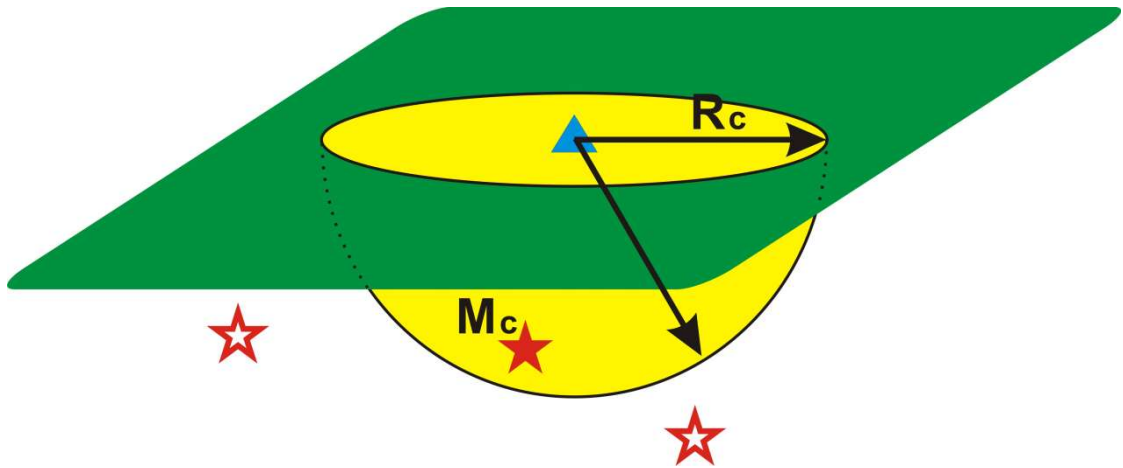

(b)

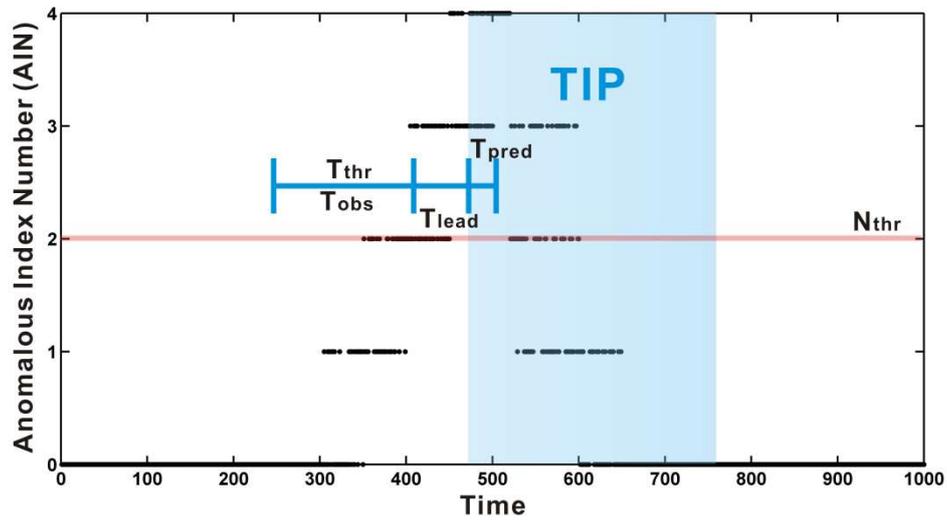

Figure 8. Schematic diagrams of (a) selection of target earthquakes, and (b) definition of 'Time of Increased Probability (*TIP*).' The filled pentagram is the target event with a source-to-station distance smaller than or equal to $R_c$ and with magnitude greater than or equal to $M_c$. An anomalous day is defined when the quantity *AIN* is greater than or equal to $N_{thr}$. The *TIPs* (blue region) within $T_{pred}$ days are issued when the number of anomalous days greater than or equal to $T_{thr}$ within $T_{obs}$ days. The leading time $T_{lead}$ accounts for the fact that an event tends to occur some finite time after the last anomaly.

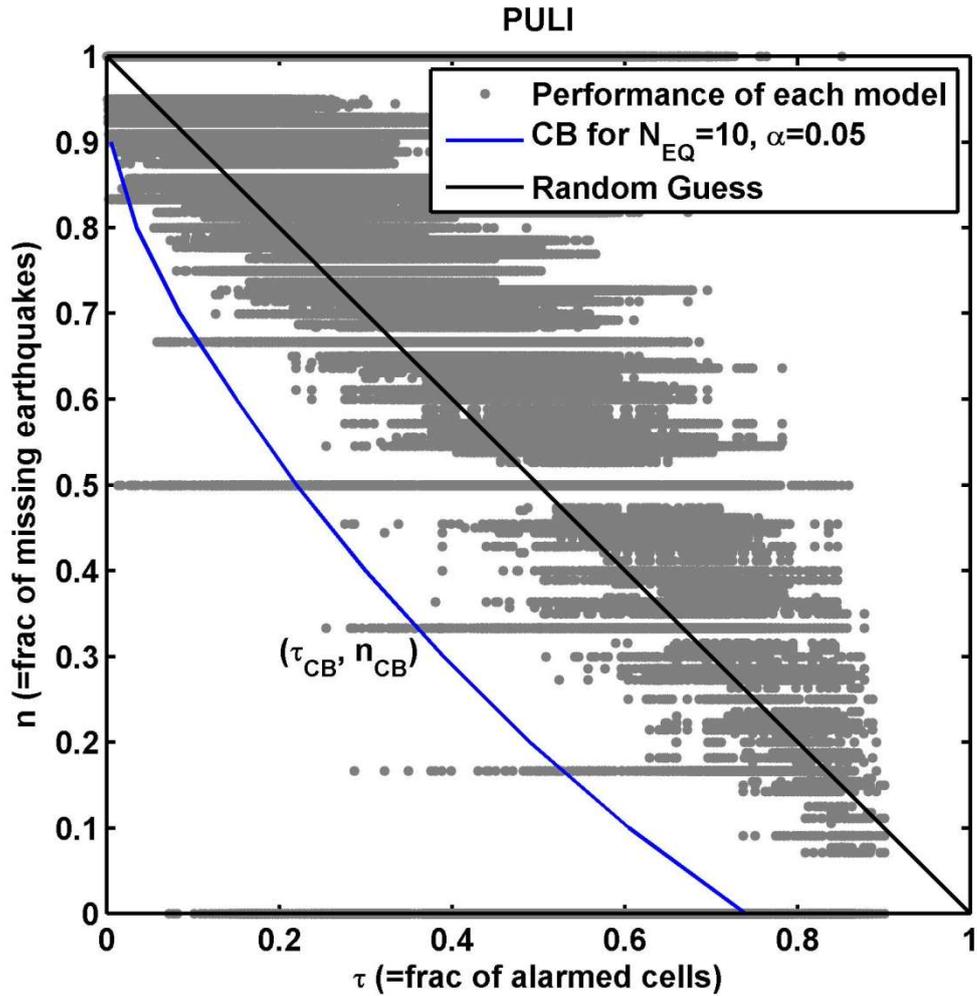

Figure 9. Molchan error diagram of the PULI station (gray dots) for the training datasets from its onset to 2015/03/31 with a confidence bound (blue line). This bound is defined at significance level $\alpha=0.05$, and the number of target events $N_{EQ}=10$. The black anti-diagonal line corresponds to random guesses.

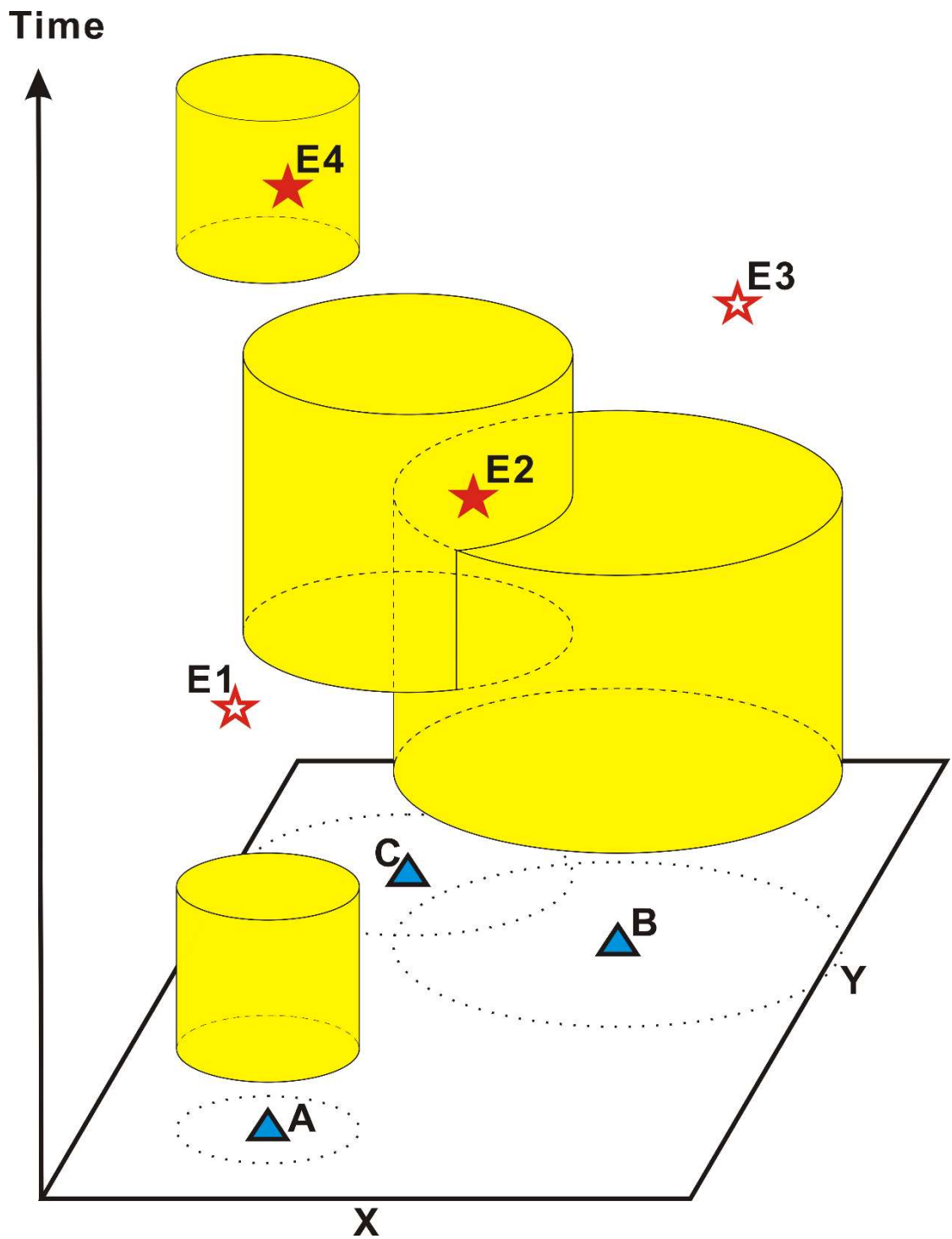

Figure 10. Schematic diagram of the joint stations method of the GEMSTIP algorithm. Triangles stand for stations, yellow regions for predicted spatio-temporal regions, open pentagrams for missed events, and filled pentagrams for hit events.

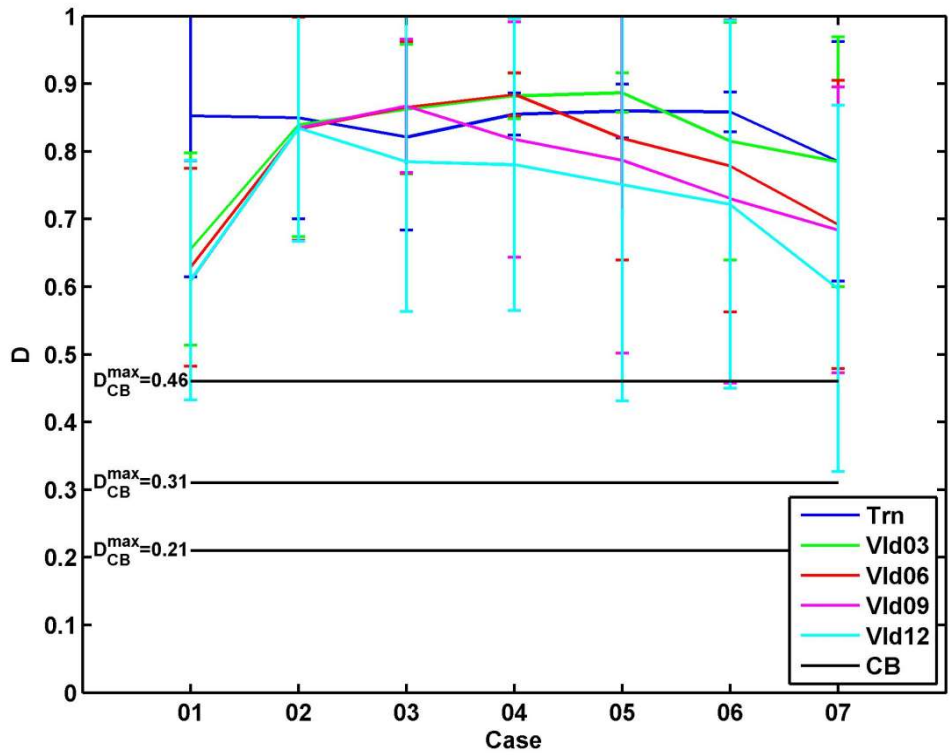

Figure 11. *D* scores (mean ± 2 standard deviations) versus cases 01-07 defined in Table 3. The black lines stand for the maximum confidence bounds $D_{cb}^{max} = 0.46$ for *α*=0.05 and *N*$_{EQ}$=5, $D_{cb}^{max} = 0.31$ for *α*=0.05 and *N*$_{EQ}$=10, and $D_{cb}^{max} = 0.21$ for *α*=0.05 and *N*$_{EQ}$=20.

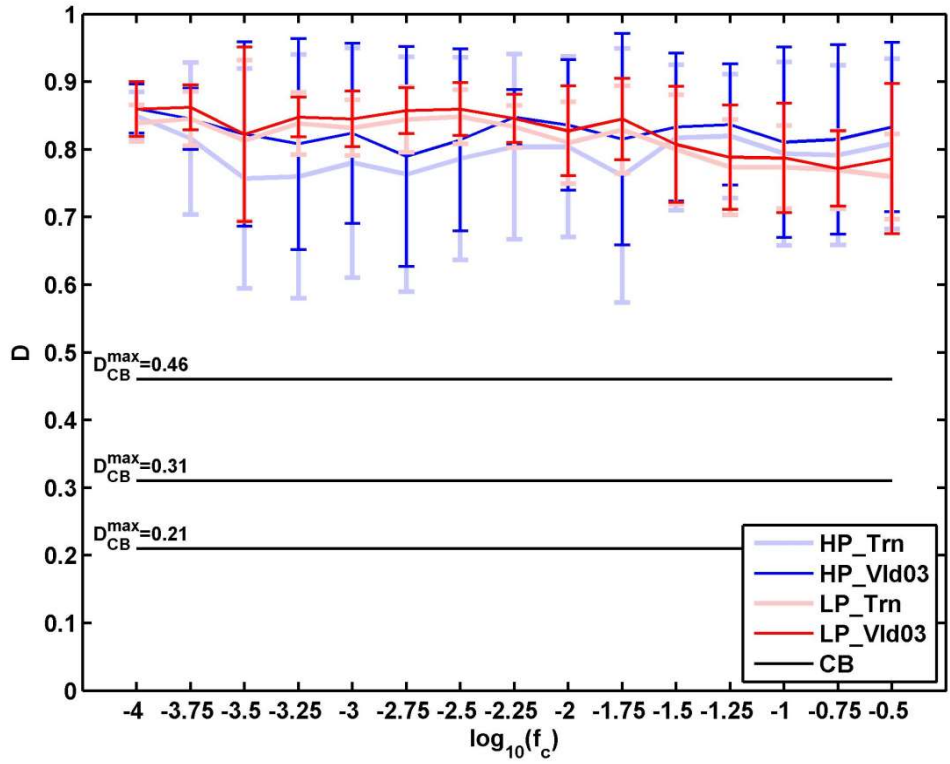

Figure 12. *D* scores (mean ± 2 standard deviations) for the low- and high-pass filtered datasets versus cut-off frequencies $f_c$ (in Hz). The blue lines represent the results of the high-pass filtered datasets, while the red ones stand for the low-pass filtered datasets. The light colors show the results of the training datasets from the onset time of each station up to 2015/03/31, while the dark colors stand for the validation datasets from 2015/04/01 to 2015/06/30.

(a)

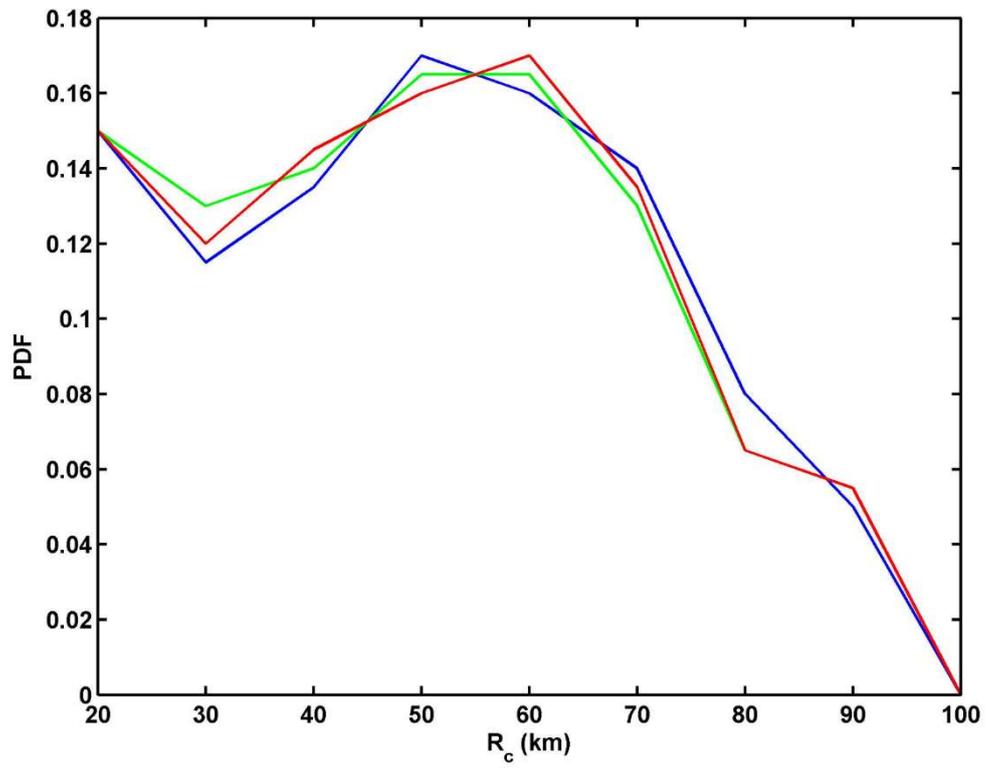

(b)

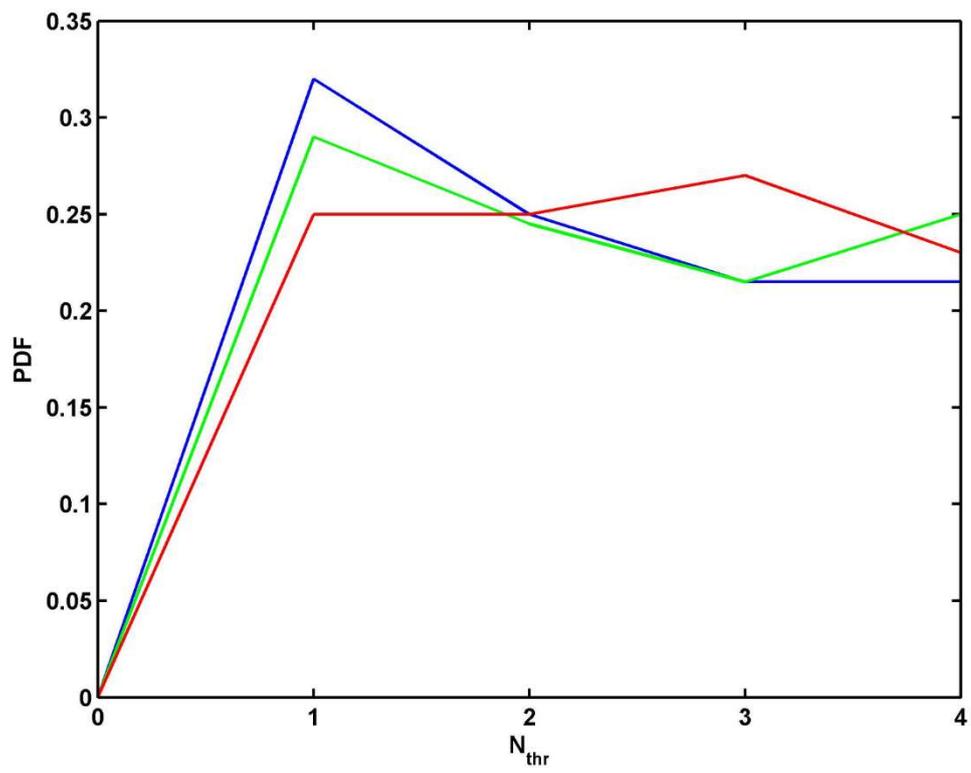

(c)

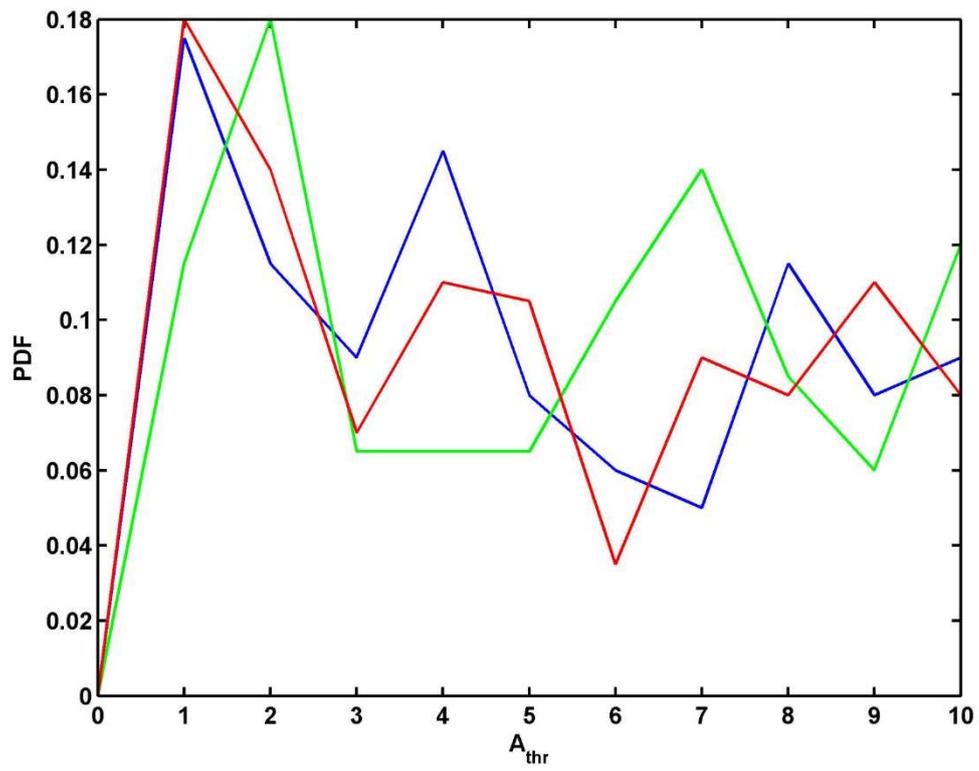

(d)

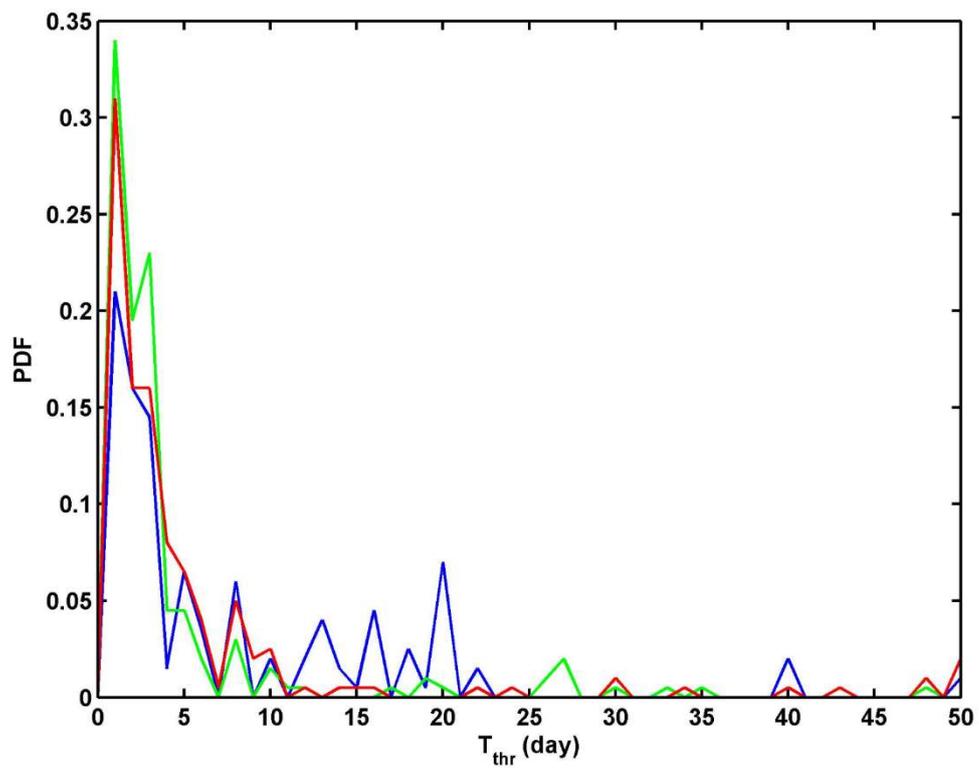

(e)

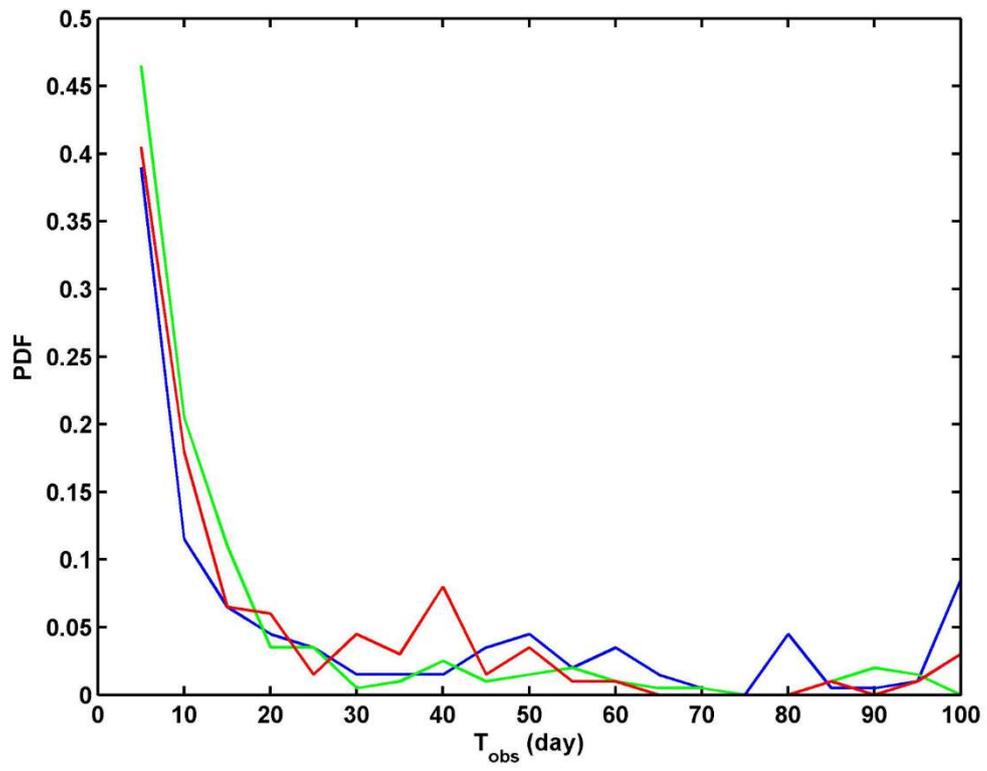

(f)

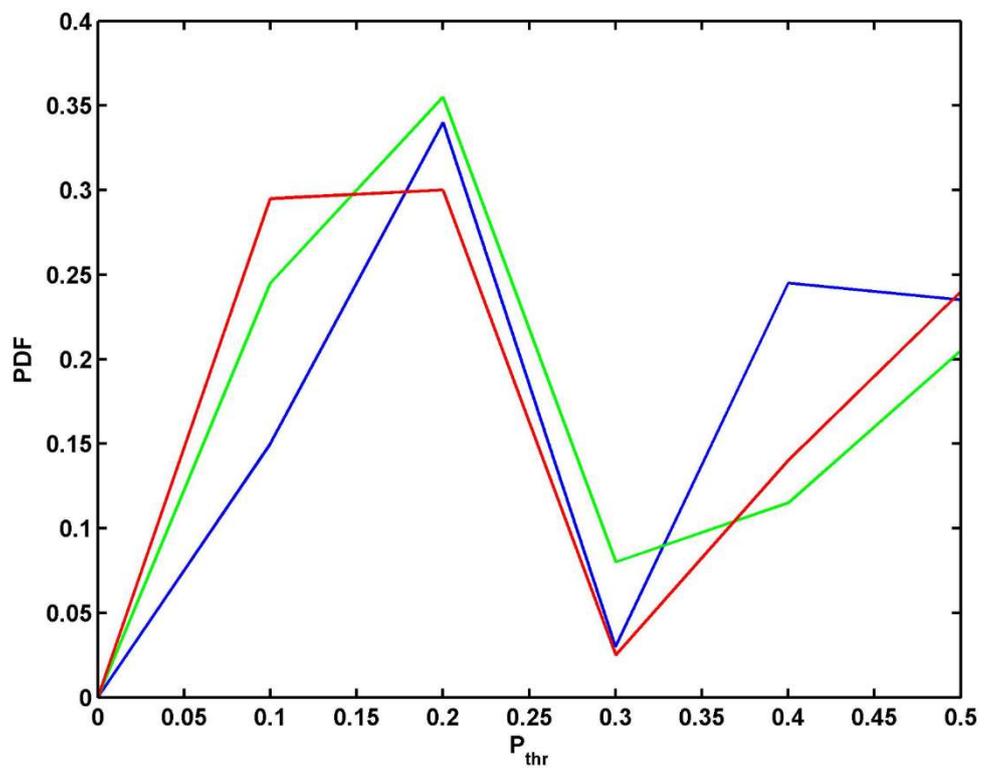

(g)

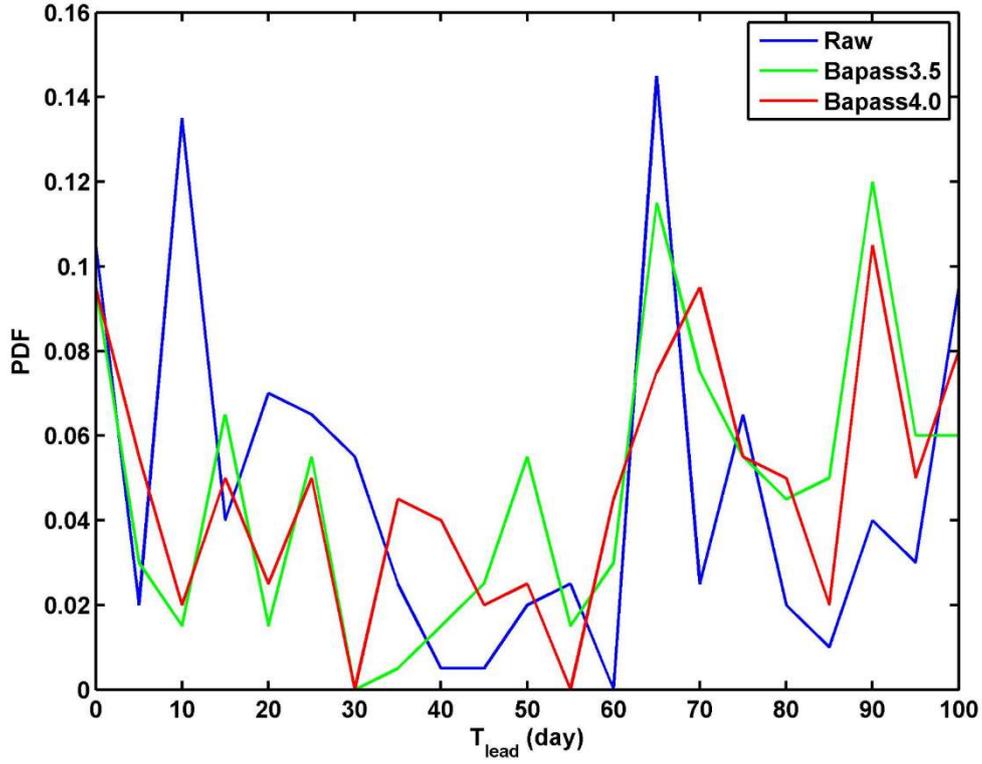

Figure 13. Probability density functions (PDFs) of the GEMSTIP parameters $(R_c, N_{thr}, A_{thr}, T_{thr}, T_{obs}, P_{thr}, T_{lead})$ shown in panels (a) to (g), respectively. Note that $P_{thr} = \frac{T_{thr}}{T_{obs}}$. These PDFs are derived from the top 10 model parameters for all stations (see Tables S1, S2, and S3 in detail). Blue lines stand for the raw dataset, green lines for the Bapass3.5 dataset, and red lines for the Bapass4.0 dataset. Note that $M_c$=5 and $T_{pred}$=1.

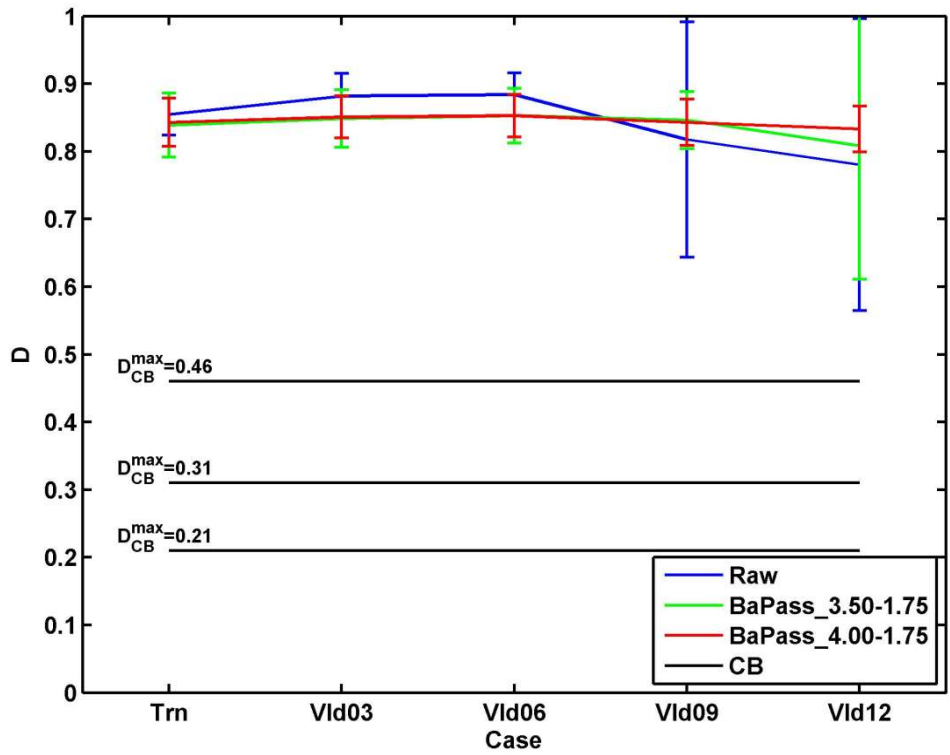

Figure 14. *D* scores (mean ± 2 standard deviations) for the raw data (blue line) and two band-pass filtered datasets. One frequency band is a control group from $10^{-3.5}$ to $10^{-1.75}$ Hz (green line), and the other is the optimal band from $10^{-4.0}$ to $10^{-1.75}$ Hz (red line). The training phase is from the onset time of each station up to 2015/03/31 (denoted as Trn in the x-axis). The validation phases are 3, 6, 9, and 12 months following the training phase (denoted as Vld03, Vld06, Vld09, and Vld12, respectively).

(a)

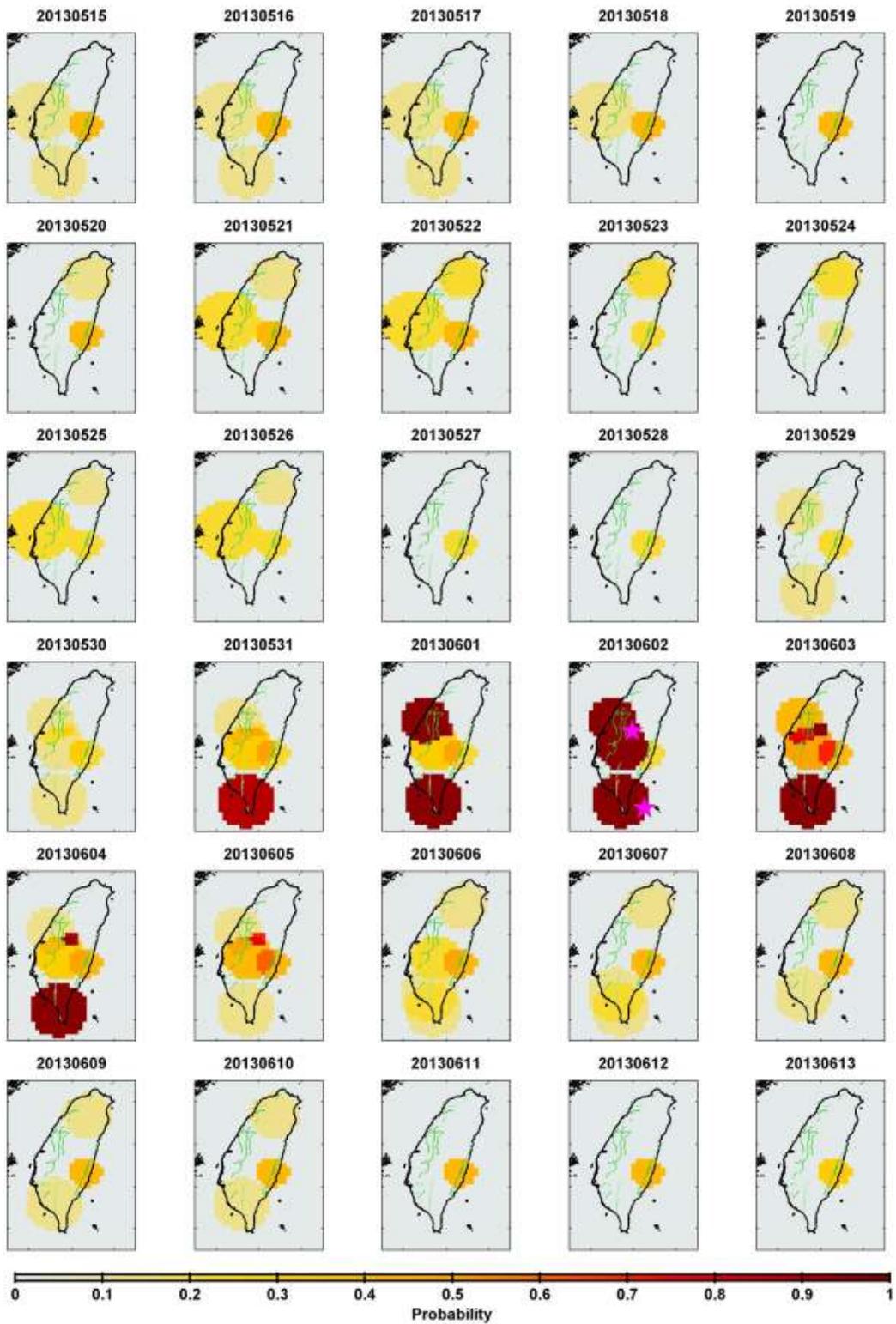

(b)

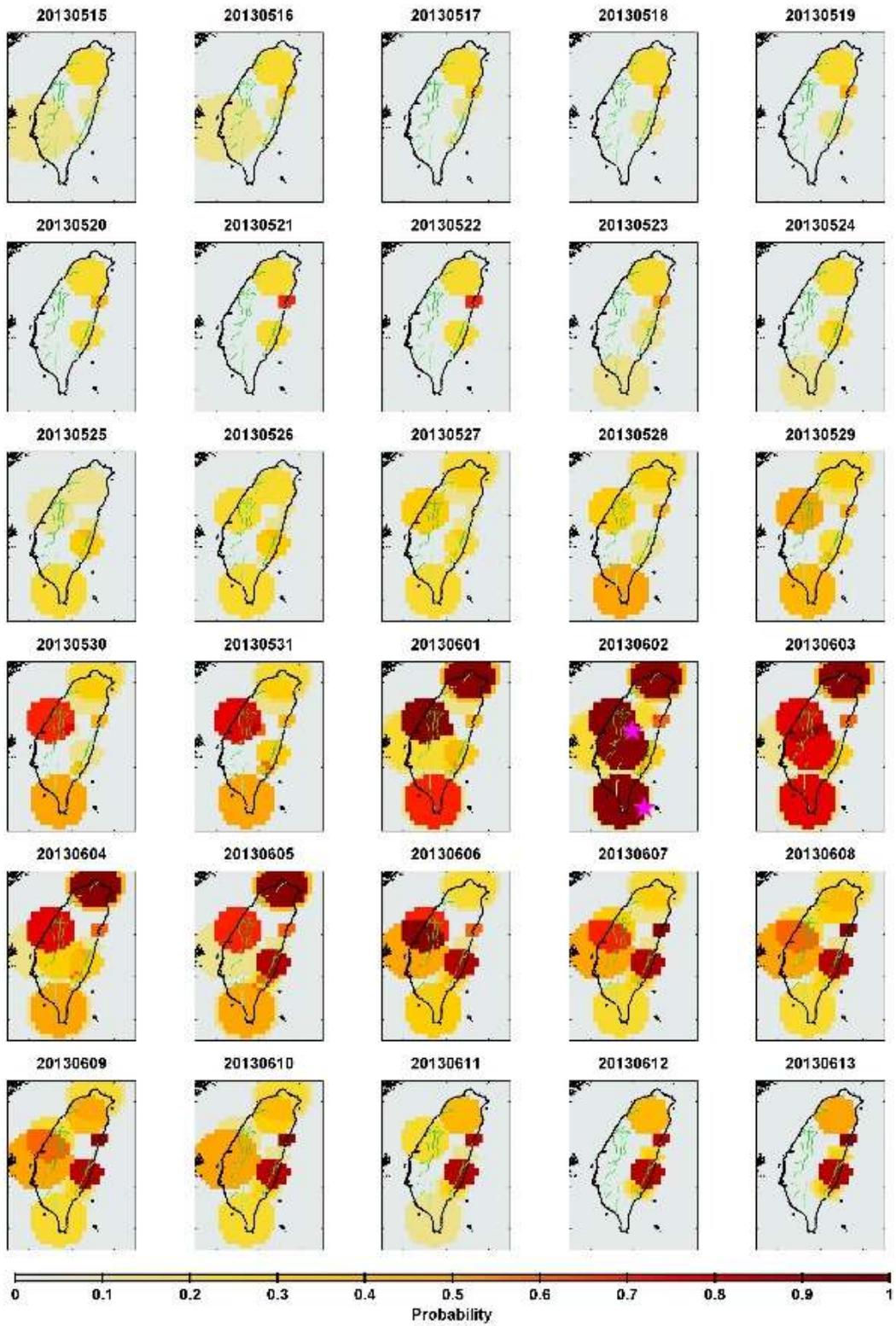

(c)

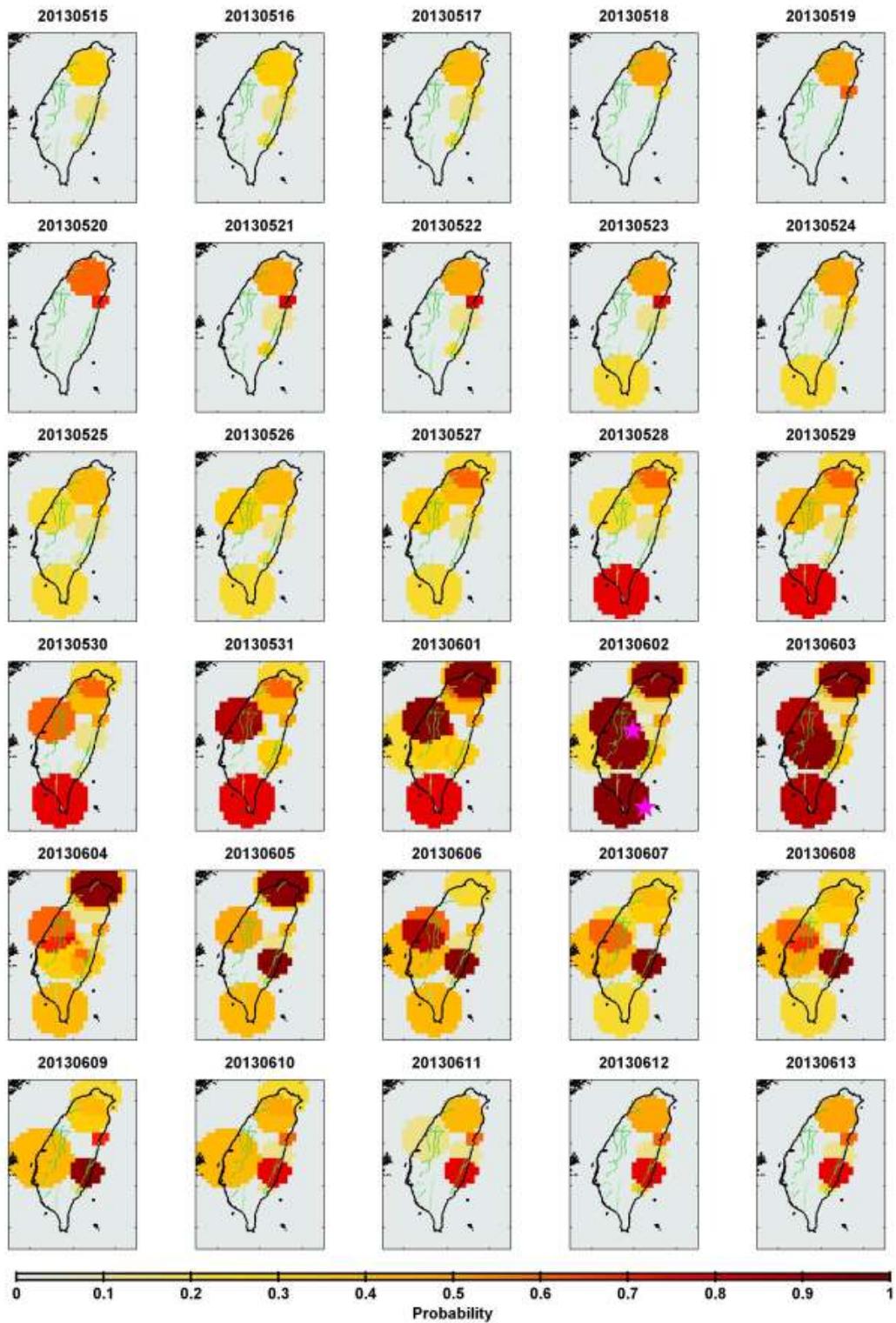

Figure 15. Spatio-temporal earthquake probability forecasts. Figures (a) to (c) stand for the raw, Bapass3.5, and Bapass4.0 datasets, respectively.

# Tables and table captions

Table 1. Information about the geoelectric stations.

| Station Name | Longitude (°E) | Latitude (°N) | Dipole length of N-component (km) | Dipole length of E-component (km) | Azimuth of N-component | Azimuth of E-component | Onset time (yyyy.mm.dd) |
|---|---|---|---|---|---|---|---|
| SHRL | 121.5619 | 25.1559 | 1.79 | 2.72 | 45.6 | 121.2 | 2012.05.18 |
| KUOL | 121.1420 | 24.9629 | 1.89 | 3.57 | 15.0 | 96.0 | 2011.10.01 |
| TOCH | 121.8052 | 24.8435 | 2.58 | 3.09 | -37.6 | 46.9 | 2012.02.10 |
| HUAL | 121.3677 | 24.6745 | 4.29 | 1.83 | -9.8 | 74.7 | 2012.01.07 |
| ENAN | 121.7849 | 24.4758 | 0.99 | 1.93 | 54.0 | 123.0 | 2012.02.15 |
| DAHU | 120.9024 | 24.4106 | 2.56 | 4.65 | -47.1 | 48.9 | 2012.02.07 |
| LISH | 121.2524 | 24.2495 | 0.61 | 0.91 | 54.0 | 100.0 | 2012.11.07 |
| SHCH | 121.6250 | 24.1183 | 3.17 | 2.99 | 9.8 | 72.4 | 2012.04.25 |
| HERM | 120.5015 | 24.1088 | 1.69 | 0.66 | -38.1 | 119.3 | 2012.02.09 |
| PULI | 120.9788 | 23.9208 | 1.63 | 2.24 | 44.3 | 145.5 | 2012.03.01 |
| FENL | 121.4112 | 23.7156 | 1.15 | 1.31 | -9.6 | 100.9 | 2012.06.28 |
| SIHU | 120.2293 | 23.6370 | 2.84 | 2.94 | 21.1 | 96.9 | 2012.02.08 |
| DABA | 120.7494 | 23.4544 | 0.29 | 0.36 | -45.7 | 48.0 | 2013.02.01 |
| YULI | 121.3181 | 23.3247 | 6.14 | 2.20 | 10.7 | 100.6 | 2012.04.24 |
| CHCH | 120.1618 | 23.2197 | 2.35 | 2.69 | 4.3 | 79.3 | 2012.05.04 |
| LIOQ | 120.6632 | 23.0321 | 0.86 | 0.51 | 37.4 | 107.5 | 2012.07.20 |
| RUEY | 121.1557 | 22.9732 | 1.67 | 1.98 | 0.8 | 106.7 | 2012.04.23 |
| KAOH | 120.2893 | 22.6577 | 1.76 | 2.02 | 21.2 | 87.6 | 2012.03.22 |
| WANL | 120.5937 | 22.5909 | 2.05 | 2.12 | -3.8 | 87.8 | 2012.12.20 |
| FENG | 120.7007 | 22.2043 | 1.76 | 0.80 | 26.8 | 130.2 | 2013.01.03 |

*Azimuth of 0 degree is the exact north direction; a positive value means clockwise rotation.

Table 2. Parameters and their value ranges for the GEMSTIP model.

| Parameter | Value |
|---|---|
| $M_c$ | 5 |
| $R_c$ | 20-100 (km) |
| $A_{thr}$ | 1-10 |
| $N_{thr}$ | 1-4 |
| $P_{thr}$ | 0.1-0.5 |
| $T_{thr}$ | $\lceil P_{thr} * T_{obs} \rceil$ (day) |
| $T_{obs}$ | 5-100 (day) |
| $T_{lead}$ | 0-100 (day) |
| $T_{pred}$ | 1 (day) |

*$\lceil x \rceil$ means the ceiling of $x$, i.e. the smallest integer greater than or equal to $x$.

Table 3. Periods, scores, and the number of earthquakes for different training sets and their corresponding validation sets.

| Case | Training Phase (Trn) | | Validation Phase of 3 months (Vld03) | | Validation Phase of 6 months (Vld06) | | Validation Phase of 9 months (Vld09) | | Validation Phase of 12 months (Vld12) | |
|---|---|---|---|---|---|---|---|---|---|---|
| | Time | ($\tau$, $n$, $D$) | Time | ($\tau$, $n$, $D$) | Time | ($\tau$, $n$, $D$) | Time | ($\tau$, $n$, $D$) | Time | ($\tau$, $n$, $D$) |
| 01 | sta. onset-2014/6/30 (47) | (0.10±0.04, 0.05±0.22, 0.85±0.24) | 2014/7/1-2014/9/30 (4) | (0.11±0.04, 0.24±0.10, 0.66±0.14) | 2014/7/1-2014/12/31 (8) | (0.13±0.06, 0.24±0.10, 0.63±0.14) | 2014/7/1-2015/3/31 (14) | (0.14±0.06, 0.25±0.12, 0.61±0.18) | 2014/7/1-2015/6/30 (19) | (0.14±0.06, 0.25±0.12, 0.61±0.18) |
| 02 | sta. onset-2014/9/30 (51) | (0.12±0.04, 0.03±0.14, 0.85±0.14) | 2014/10/1-2014/12/31 (4) | (0.13±0.04, 0.03±0.14, 0.84±0.16) | 2014/10/1-2015/3/31 (10) | (0.13±0.04, 0.03±0.14, 0.83±0.16) | 2014/10/1-2015/6/30 (15) | (0.13±0.04, 0.03±0.14, 0.83±0.16) | 2014/10/1-2015/9/30 (20) | (0.13±0.04, 0.03±0.14, 0.83±0.16) |
| 03 | sta. onset - 2014/12/31 (55) | (0.15±0.04, 0.03±0.14, 0.82±0.14) | 2015/1/1-2015/3/31 (6) | (0.12±0.02, 0.02±0.10, 0.86±0.10) | 2015/1/1-2015/6/30 (11) | (0.12±0.02, 0.02±0.10, 0.87±0.10) | 2015/1/1-2015/9/30 (16) | (0.12±0.02, 0.02±0.10, 0.87±0.10) | 2015/1/1-2015/12/31 (22) | (0.13±0.02, 0.09±0.22, 0.78±0.22) |
| 04 | sta. onset - | (0.15±0.04, | 2015/4/1- | (0.12±0.04, | 2015/4/1- | (0.12±0.04, | 2015/4/1- | (0.13±0.04, | 2015/4/1- | (0.15±0.04, |

| | | | | | | | | | | | | |
|---|---|---|---|---|---|---|---|---|---|---|---|---|
| | | 2015/3/31 (61) | (0.00±0.00, 0.85±0.04) | 2015/6/30 (5) | (0.00±0.00, 0.88±0.04) | 2015/9/30 (10) | (0.00±0.00, 0.88±0.04) | 2015/12/31 (16) | (0.05±0.18, 0.82±0.18) | 2016/3/31 (23) | (0.07±0.22, 0.78±0.22) |
| 05 | sta. onset - 2015/6/30 (66) | (0.14±0.04, 0.00±0.00, 0.86±0.04) | 2015/7/1-2015/9/30 (5) | (0.11±0.02, 0.00±0.00, 0.89±0.02) | 2015/7/1-2015/12/31 (11) | (0.13±0.04, 0.05±0.18, 0.82±0.18) | 2015/7/1-2016/3/31 (18) | (0.15±0.04, 0.07±0.28, 0.79±0.28) | 2015/7/1-2016/6/30 (30) | (0.15±0.04, 0.10±0.30, 0.75±0.32) | | |
| 06 | sta. onset - 2015/9/30 (71) | (0.14±0.02, 0.00±0.00, 0.86±0.02) | 2015/10/1-2015/12/31 (6) | (0.13±0.04, 0.05±0.18, 0.82±0.18) | 2015/10/1-2016/3/31 (13) | (0.15±0.04, 0.07±0.22, 0.78±0.22) | 2015/10/1-2016/6/30 (25) | (0.16±0.04, 0.11±0.26, 0.73±0.28) | 2015/10/1-2016/9/30 (25) | (0.16±0.04, 0.11±0.26, 0.72±0.28) | | |
| 07 | sta. onset - 2015/12/31 (77) | (0.16±0.04, 0.05±0.16, 0.79±0.18) | 2016/1/1-2016/3/31 (7) | (0.16±0.02, 0.05±0.20, 0.78±0.18) | 2016/1/1-2016/6/30 (19) | (0.17±0.04, 0.14±0.22, 0.69±0.22) | 2016/1/1-2016/9/30 (19) | (0.18±0.04, 0.14±0.22, 0.68±0.22) | 2016/1/1-2016/12/31 (28) | (0.19±0.04, 0.22±0.28, 0.60±0.28) | | |

*The ($\tau$, $n$, $D$) are represented by mean ± 2 standard deviations, and rounded to hundredths. The figure in a bracket of the Time column is the number of earthquakes during that time period.

Table 4. Scores, and the number of earthquakes for the training sets from the station onset time to 2015/3/31 and its corresponding validation sets for two band-pass filtered datasets.

| Dataset | Training Phase (Trn) | | Validation Phase of 3 months (Vld03) | | Validation Phase of 6 months (Vld06) | | Validation Phase of 9 months (Vld09) | | Validation Phase of 12 months (Vld12) | |
|---|---|---|---|---|---|---|---|---|---|---|
| | Time | ($\tau$, $n$, $D$) | Time | ($\tau$, $n$, $D$) | Time | ($\tau$, $n$, $D$) | Time | ($\tau$, $n$, $D$) | Time | ($\tau$, $n$, $D$) |
| Bapass 3.5 | sta. onset - 2015/3/31 (61) | (0.16±0.04, 0.00±0.00, 0.84±0.04) | 2015/4/1- 2015/6/30 (5) | (0.15±0.04, 0.00±0.00, 0.85±0.04) | 2015/4/1- 2015/9/30 (10) | (0.15±0.04, 0.00±0.00, 0.85±0.04) | 2015/4/1- 2015/12/31 (16) | (0.15±0.04, 0.00±0.00, 0.85±0.04) | 2015/4/1- 2016/3/31 (23) | (0.16±0.04, 0.03±0.18, 0.81±0.20) |
| Bapass 4.0 | sta. onset - 2015/3/31 (61) | (0.16±0.04, 0.00±0.00, 0.84±0.04) | 2015/4/1- 2015/6/30 (5) | (0.15±0.04, 0.00±0.00, 0.85±0.04) | 2015/4/1- 2015/9/30 (10) | (0.15±0.04, 0.00±0.00, 0.85±0.04) | 2015/4/1- 2015/12/31 (16) | (0.16±0.04, 0.00±0.00, 0.84±0.04) | 2015/4/1- 2016/3/31 (23) | (0.17±0.04, 0.00±0.00, 0.83±0.04) |

*The ($\tau$, $n$, $D$) are represented by mean ± 2 standard deviations, and rounded to hundredths. The figure in a bracket of the Time column is the number of earthquakes during that time period. Bapass3.5 dataset stands for filtering with frequency from $10^{-3.5}$ to $10^{-1.75}$ Hz, and Bapass4.0 for filtering with frequency from $10^{-4.0}$ to $10^{-1.75}$ Hz.